\documentclass[preprint,groupedaddress,floatfix,showpacs,showkeys]{revtex4-1}
\usepackage{hyperref}
\usepackage{amsmath}
\usepackage{amssymb}
\usepackage{textcomp}
\usepackage[mathlines]{lineno}
\usepackage{graphicx}
\usepackage[american]{babel}
\usepackage{afterpage}
\usepackage{grffile} 
\usepackage{siunitx} 
\usepackage{multirow}
\usepackage{subcaption} 
\newcommand{\vphi}{\mbox{\boldmath{$\phi$}}}

\newcommand{\vmu}{\mbox{\boldmath{$\mu$}}}
\usepackage[toc,page]{appendix}
\usepackage{adjustbox}
\usepackage{array}
\usepackage{bm}
\usepackage{xcolor}
\usepackage{float}
\begin{document}
\raggedbottom

\title{Role of interfacial energy and liquid diffusivities on pattern formation during thin-film three-phase eutectic growth}
\author{Sumeet Khanna}
\email{sumeet92k@gmail.com}
\affiliation{Department of Materials Engineering, Indian Institute of Science, Bengaluru, India, Pin: 560012}
\author{Abhik Choudhury}
\email{Corresponding author: abhiknc@iisc.ac.in}
\affiliation{Department of Materials Engineering, Indian Institute of Science, Bengaluru, India, Pin: 560012}
\begin{abstract}
In this paper, we investigate three-phase eutectic growth during thin-film directional solidification of a model symmetric ternary eutectic alloy. In contrast to two-phase eutectics that have only a single possibility, $\alpha\beta\alpha\beta\ldots$, as the growth pattern, during three-phase eutectic growth infinite possibilities exist. Here, we explore the possible existence of pattern selection influenced by the change in the solid-solid interfacial energies and the diffusivities. We begin the study by estimating the undercooling vs. spacing variation for the simplest possible configurations of pattern lengths 3 and 4, where phase-field simulations are utilized to quantify the influence of the solid-solid interfacial energy and the contrast in the component diffusivities. Subsequently, extended simulations consisting of multiple periods of $\alpha\beta\delta\ldots$ as well as $\alpha\beta\delta\beta\ldots$ are carried out for assessing the stability of the configurations to long-wavelength perturbations in spacing. Thereafter, growth competition among the simplest patterns is investigated through phase-field simulations of coupled growth of configurations of the type $\left[\alpha\beta\delta\right]_m\left[\alpha\beta\delta\beta\right]_n$, (m,n) being the respective number of periods. Finally, pattern selection is studied by initializing with random initial configurations and classifying the emerging patterns based on the phase sequences. 
The principal finding is that, while there is no strong phase selection, between the solid-solid interfacial energies and the contrast in the component diffusivities, we find the latter to strongly influence pattern morphology.

\end{abstract}

\maketitle
\section{Introduction}
Coupled growth of eutectic alloys presents an interesting problem for both physicists and material scientists. From a theoretical perspective, understanding the growth and dynamics of these systems is crucial to understanding pattern formation in non-equilibrium systems. For a metallurgist, the coupled growth presents a vast combination of multiphase microstructures, which provide avenues for tweaking the mechanical, chemical and functional properties of materials. 
Although the interest of the metallurgical industry is in bulk eutectic microstructures, from a scientific standpoint, a deeper understanding of microstructure evolution during thin sample growth becomes useful for the investigation of even more complex dynamics of pattern formation during bulk solidification.
In this paper, we explore 2D lamellar growth in ternary three-phase eutectics using the phase-field method under directional solidification conditions.

We begin with a brief review of the studies of eutectic growth and an extended survey can be found in \cite{akamatsu2016eutectic}.
The first studies on the coupled growth of phases during eutectic/eutectoid reactions are due to Zener \cite{zener1946kinetics}, that are subsequently improved and extended by Tiller  \cite{tiller1958liquid}, Hillert \cite{hillert1960structure}, and Jackson and Hunt (JH) \cite{jh1966} which is the most widely used theory for the investigation of eutectic microstructure evolution.
The JH theory presents a simplified version of the complex Stefan-problem of two-phase eutectic growth leading to the solution of the interface shape, the composition fields as well as the variation of the solidification front undercooling as a function of spacing. 
The operating point in eutectic morphologies is derived as the minimum undercooling spacing $\lambda_{min}$ with the scaling relation $\lambda_{min}^2v=$ constant, ($v$ being the velocity). 
The simplicity of the calculations lends to its wide utilization to predict the scale of the microstructure in experiments. Langer \cite{langer1980eutectic} provides a theory for the temporal evolution of spacings in response to long-wavelength perturbations through which he presents a quantitative rationale for $\lambda_{min}$ being the marginal stability point for eutectic growth, where spacings lower than $\lambda_{min}$
are susceptible to elimination (also referred to as the Eckhaus instability) and topological mechanisms/instabilities (in bulk solidification conditions) exist for reduction in spacings larger than the $\lambda_{min}$. While $\lambda_{min}$ is therefore a reasonable approximation for the observed eutectic microstructural scale, in reality there exists a small range of stable spacings around $\lambda_{min}$, where spacings lower than $\lambda_{min}$ could also be stable to the Eckhaus instability \cite{akamatsu2002pattern,akamatsu2004overstability}. 
This experimental observation has also been confirmed through phase-field simulations and the difference with the predictions from Langer's theory arises because of the assumption that the motion of the tri-junction is normal to the local solidification envelope, which is shown to be incorrect from the observations of tri-junction motion in phase-field simulations \cite{plapp2002eutectic, akamatsu2002pattern, akamatsu2004overstability}.

Spacings larger than $\lambda_{min}$ in thin-film solidification conditions are prone to oscillatory and tilt instabilities. The existence of oscillatory instabilities beyond a critical spacing is first shown by  Datye and Langer \cite{datye1981stability}. Further, mechanisms of symmetry breaking include tilt instabilities that have been studied theoretically and experimentally \cite{kassner1991spontaneous, faivre1992tilt}.
Later, Karma and Sarkissian \cite{karma1996morphological} utilize the boundary element method for computing stability diagrams and estimate the complete range of instabilities, that are in good agreement with experimental findings \cite{ginibre1997experimental}.  For bulk solidification conditions, the zig-zag instability has been simulated using the phase-field model \cite{parisi2008stability}. 

Similar to binary eutectic alloys several experimental studies have been conducted on three-phase eutectic alloys, that include Ag-Al-Cu \cite{mccartney1980structures,mccartney1980structures2,boyuk2009directional, genau2012morphological, dennstedt2012microstructures,dennstedt20163d, steinmetz2018study}, In-Bi-Sn \cite{ruggiero1997origin, apel2004lamellar, rex2005transient, witusiewicz2005situ, bottin2016stability, mohagheghi2017dynamics, mohagheghi2018quasi, mohagheghi2019effects}, Sn-Ag-Cu \cite{boyuk2009microstructure}, as well as a number of transparent organic systems \cite{sturz2004organic, witusiewicz2006phase}.
Here, we will focus on the issues related to thin-film solidification where in contrast to two-phase growth there exist infinite possibilities for the growth configurations (e.g., $\alpha\beta\delta\ldots$, $\alpha\beta\delta\beta\ldots$ and so on). Mirroring theoretical efforts on binary eutectic two-phase growth, derivations similar to the JH theory for predicting undercooling vs. spacing variations are performed by Himemiya et al. \cite{himemiya1999three} for certain lamellar configurations, and a more generic derivation for any periodic arrangement of lamellae is available in Choudhury et al. \cite{choudhury2011theoretical}. 
Further in \cite{choudhury2011theoretical} a comparison is made between the undercooling vs. spacing relationships derived analytically and that measured from phase-field simulations where a symmetric ternary eutectic alloy has been considered. Additionally, the different modes of oscillatory instabilities have been investigated using phase-field simulations and the symmetry-breaking modes are compared with those observed in the case of two-phase eutectic growth. A novel short-wavelength instability is also derived which leads to the transformation of a single period of $\alpha\beta\delta\beta$ to $\alpha\beta\delta$ below a critical spacing. 

With regards to pattern selection, experimental and modeling studies indicate the formation of patterns that are mirror symmetric, for example  $\alpha\beta\alpha\delta$ \cite{apel2004lamellar}. With the development of novel quasi-2D directional solidification techniques and in-situ characterization, experimentalists have studied the long-wavelength perturbations in lamellar spacing and stability limits of the $\alpha \beta \alpha \delta$ pattern in the In-Bi-Sn eutectic alloy using controlled directional solidification experiments \cite{mohagheghi2017dynamics, bottin2016stability}. Additionally, the coexistence of different lamellar patterns with larger periodic cycles like the $[\alpha \beta]_m[\alpha \delta]_n$ along with the $\alpha \beta \alpha \delta$ pattern are reported and transitioning mechanisms between patterns have been suggested. However, the parameters influencing the pattern selection are yet unclear. For asymmetric patterns such as $\alpha\beta\delta$ which occur as transients during the solidification process, tilts of the solid-solid interface with respect to the temperature gradient direction in directional solidification is found to occur due to the absence of mirror symmetry \cite{bottin2016stability} that have also been observed in phase-field simulations \cite{hotzer2016phase} for unequal interfacial energies as well as due to contrasting component diffusivities \cite{lahiri2017revisiting} and asymmetric concentration profiles \cite{hecht2004multiphase,apel2004lamellar} brought about by unequal phase fractions or asymmetry in the phase diagram. 
While in the In-Bi-Sn eutectic it has been observed that the short $\alpha \beta \delta$ transients that grow tilted are eliminated to form the $\alpha \beta \delta \beta$ pattern \cite{bottin2016stability}, a modeling study that investigates the competing growth of different patterns and quantitatively assesses why some patterns are observed more than others is lacking.

This forms the motivation for this paper, where we will utilize phase-field simulations to investigate pattern competition and pattern selection upon variation of two principal parameters; solid-solid interfacial energies and the solutal diffusivity contrast while keeping the phase-diagram symmetric. While the number of parameters is vast, the mentioned properties bring about changes in the two factors influencing eutectic growth, the capillarity and the diffusion and therefore we expect to derive broad insights with wide applicability across different ternary eutectic systems.
The studies will first determine the influence of contrast in solutal diffusivities and difference in solid-solid interfacial energies among the three interfaces on the undercooling vs. spacing variation of the simplest configurations i.e $\alpha\beta\delta\ldots$ and  $\alpha\beta\delta\beta\ldots$ (along with permutations) in Section \ref{sec:constrained}. Subsequently, temporal evolution of spacing in the simplest configurations, in response to long-wavelength perturbations will be studied in Section \ref{sec:stability}. Thereafter, in Section \ref{sec:competition}, competing simulations involving multiple periods of the simplest configurations will be conducted for varying interfacial energies as well as solutal diffusivity contrast. Finally in Section \ref{sec:complex}, simulations starting from random initial configurations are performed and the steady state patterns are classified by the phase sequences.
The simulation studies will reveal the transformation mechanisms between the different configurations as well as the influence of the asymmetries with respect to the interfacial energies and the diffusivities on the evolution of steady state patterns.

\section{Methodology}
\label{sec:methodology}
We use a multi phase-field model based upon the grand potential formulation \cite{choudhury2012grand,plapp2011unified} and applied
in microstructure simulations of three-phase eutectic growth in \cite{choudhury2015pattern, bauer2015massively,choudhury2016quantification,steinmetz2016large, lahiri2017revisiting}.
The volume fraction of each of the N phases (N = 4) is represented as $\left(\phi_\alpha, \phi_\beta, \phi_\delta, \phi_l\right)$, and the phase evolution equation for each phase is written as, 
\begin{flalign}
\tau \epsilon \frac{\partial \phi_{\alpha}}{\partial t}=& \epsilon\left(\nabla \cdot \frac{\partial a(\phi, \nabla \phi)}{\partial \nabla \phi_{\alpha}}-\frac{\partial a(\phi, \nabla \phi)}{\partial \phi_{\alpha}}\right) 
-\frac{1}{\epsilon} \frac{\partial w(\phi)}{\partial \phi_{\alpha}}-\frac{\partial \psi(T, \mu, \phi)}{\partial \phi_{\alpha}}-\lambda_V,
\label{phi_eqn}
\end{flalign}
where $\tau$ is the relaxation constant that controls the interface kinetics \cite{choudhury2012grand} and is calculated as in \cite{choudhurythesis, khanna2020role}. 
$\epsilon$ controls the diffuse interface width.  $a(\phi, \nabla \phi)$ is the gradient energy term calculated as $a(\phi, \nabla \phi) = \sum_{\alpha < \beta}^{N, N} \gamma^{\alpha \beta} |\vec{q}_{\, \alpha \beta}|^2$, where $\gamma^{\alpha \beta}$ is the $\alpha-\beta$ interface energy and $ \vec{q}_{\, \alpha \beta} = \phi_\alpha \nabla \phi_\beta - \phi_\beta \nabla \phi_\alpha $ is the interface normal vector. 
$\psi$ is the driving force defined in terms of the grand potential, and $w(\phi)$ is the multiphase double obstacle potential.  
The Lagrange multiplier $\lambda_V$ is used to impose the constraint $\sum_\alpha^N \phi_\alpha = 1$. 
The phase-evolution equation \ref{phi_eqn} is coupled with the mass conservation equation described in terms of the diffusion potential for the (K-1=2) independent components as,
\begin{align}
\left\lbrace\dfrac{\partial \mu_i}{\partial t}\right\rbrace = &
\left[\sum_{\alpha}^N 
h^{\alpha}\left(\vphi\right)\dfrac{\partial c_i^\alpha\left(\vmu,
T\right)}{\partial \mu_j}\right]^{-1}_{ij}\Bigg\lbrace\nabla\cdot\left(\sum_{j=1}^{K-1}M_{ij}
\left(\vphi\right)\nabla \mu_j - J_{at,i}\right) \nonumber\\ 
&- \sum_{\alpha}^N c^\alpha_{i}\left(\vmu,T\right)\dfrac{\partial
h^\alpha\left(\vphi\right)}{\partial t} - \dfrac{\partial T}{\partial t}\sum_{\alpha}^N \left(\dfrac{\partial c^\alpha_{i}\left(\vmu,T\right)}{\partial T}\right)_{\vmu}
h^\alpha\left(\vphi\right)\Bigg\rbrace.
\label{eqn:mu_eqn}
\end{align}
In the preceding equation \ref{eqn:mu_eqn}, we have used the matrix-vector notation, where 
quantities in curly braces \{\} are vectors of size K-1,  while quantities in square brackets [] are matrices of size (K-1)$\times$(K-1).
The set $c^\alpha = (c_A^\alpha, c_B^\alpha)$ contains the concentrations of the independent components of each phase, $h^\alpha(\vphi)$ is the interpolation function between phases (the form is chosen the same as in \cite{khanna2020role}). 
$J_{at,i}$ is the anti-trapping current (functional form is available in \cite{choudhurythesis, khanna2020role}), and under directional solidification at temperature gradient $G_T$ and a constant velocity $v$, $\dfrac{\partial T}{ \partial t} = -G_Tv$.
The mobility matrix $M_{ij}$ is linearly interpolated between the phases with $g(\phi_\alpha) = \phi_\alpha$ as
\begin{equation}
 \left[M_{ij}\right] = \sum_\alpha^N \left[D^{\alpha}_{ik}\right] \left[\dfrac{\partial
c_k^{\alpha}\left(\vmu,T\right)}{\partial \mu_j}\right] g(\phi_\alpha).
\label{eqn:mobility}
\end{equation}
In any eutectic reaction, the diffusivity of the components in the solid phases is much lower compared to the diffusivity of the components in the liquid. Hence, we have assumed the solid phases to have zero diffusivities of the solutes, which is represented by the null matrix, i.e. a matrix having all the individual solute diffusivities, $D_{AA} = D_{BB} = D_{AB} = D_{BA} = 0$.
For the liquid phase different diagonal matrices are chosen for different conditions of diffusivities,
\begin{flalign}
\begin{bmatrix}
1 & 0\\
0 & 1
\end{bmatrix}, \quad
\begin{bmatrix}
1.6 & 0\\
0 & 1
\end{bmatrix} \quad \text{and} \quad
\begin{bmatrix}
2 & 0\\
0 & 1
\end{bmatrix}.
\end{flalign}
Thus we have three liquid diffusivity matrices for three different cases, with  increasing ratio of $D^l_{AA}/D^l_{BB}$ from 1 to 2.
For conciseness, the above three diffusivity matrices will henceforth be referred to by their diagonal components as $D^l = (1, 1)$, $D^l = (1.6, 1)$ and $D^l = (2, 1)$ respectively.
We construct a ternary phase diagram using paraboloid free energies for each of the four phases involved in the eutectic reaction. Since our principal aim is to study the influence of solid-solid interfacial energies and liquid  diffusivities on the pattern selection, we construct a symmetric phase diagram, where all the solid phases occupy equal volume fractions (1/3) at the invariant point and have symmetric liquidus and solidus slopes. As a result, the free energy paraboloid of the solid phases are symmetrically distributed around the paraboloid of the liquid phase in the centre. The free energies are represented as a function of the two independent components as \cite{choudhury2015method}:
$$ f^{\alpha}(c) = \sum_{i<j}^{2,2} A_{ij}^\alpha c_i c_j + \sum_{j}^{2} B_j^\alpha(T) c_j + E^\alpha(T), $$
where $A_{ij}^{\alpha} = A_{ij}^{l} = 1$, $B_j^l(T) = 0$ and $E^l(T) = 0$. $B_j^\alpha (T)$ and $E^\alpha (T)$ are derived for each solid phase as in \cite{choudhury2015pattern}, such that a symmetric phase diagram is produced. The thermodynamic parameters are listed in the Table \ref{table:parameters}, all in non-dimensionalized units.
The $\alpha$ phase is richer in component A, the $\beta$ phase in component B and the $\delta$ phase in component C.
\begin{table}\centering
\begin{tabular}{|c|c|c|}
\hline 
Parameter & Symbol & Value\tabularnewline
\hline 
\hline 
Eutectic Temperature & $ T_{eut} $ & 1 \tabularnewline
\hline
\multirow{4}{*}{Eutectic Composition} & $ c_{eut}^l $ & (0.333, 0.333) \tabularnewline
\cline{2-3} 
 & $ c_{eut}^{\alpha} $ & (0.6, 0.2) \tabularnewline
 \cline{2-3} 
 & $ c_{eut}^{\beta} $ & (0.2, 0.6) \tabularnewline
\cline{2-3} 
 & $ c_{eut}^{\delta} $ & (0.2, 0.2) \tabularnewline
\hline 
\multirow{3}{*}{Liquidus slope} & $ m^{l-\alpha} $ & $(1.2, 0)$ \tabularnewline
\cline{2-3} 
 & $ m^{l-\beta} $ & $(0, 1.2)$ \tabularnewline
\cline{2-3} 
 & $ m^{l-\delta} $ & $(-1.2, -1.2)$ \tabularnewline
\hline 
\multirow{3}{*}{Solidus slope} & $ m^{\alpha-l} $ & $(1.6, 0)$ \tabularnewline
\cline{2-3} 
 & $ m^{\beta-l} $ & $(0, 1.6)$ \tabularnewline
\cline{2-3} 
 & $ m^{\delta-l} $ & $(-1.6, -1.6)$ \tabularnewline
\hline 
\multirow{1}{*}{Solid-liquid interface energy} & $ \gamma^{\alpha l}, \gamma^{\beta l}, \gamma^{\delta l} $ & 0.333 \tabularnewline
\cline{2-3}
 \hline
Velocity  & $ v $  & 0.002 \tabularnewline
\hline 
Temperature gradient  & $ G_T $  & $5 \times 10^{-5}$ \tabularnewline
\hline 
\end{tabular}
\caption{Simulation parameters (non-dimensionalized units)}
\label{table:parameters}
\end{table}

In the following sections, we perform two kinds of simulations, constrained simulations and extended simulations of lamellar patterns. 
The constrained simulations are conducted to extract the solid-liquid interfacial undercooling vs. spacing relationships of the simplest lamellar patterns with the lowest number of lamellae in a period, namely the $\alpha \beta \delta$, $\alpha \beta \alpha \delta$, $\alpha \beta \delta \beta$ and $\delta \alpha \delta \beta$ lamellar patterns for different solid-solid interfacial energy configurations and liquid diffusivities, and thus provide insights about the role played by these parameters on the relative undercoolings of the different lamellar patterns.
Subsequently, in order to find out the limit of stability of the patterns to long-wavelength perturbations and also calculate the phase-diffusion coefficient,  we carry out extended simulations where the initial configuration of 10 or 20 periods of uniformly spaced patterns is slightly perturbed and the evolution dynamics is studied with time. 
In order to study the competition between the $\alpha \beta \delta$ and $\alpha \beta \delta \beta$ patterns, we conduct extended simulations starting with 5 periods of each pattern arranged as [$\alpha \beta \delta]_5[\alpha \beta \delta \beta]_5$.  These simulations are conducted for different spacings of the $\alpha \beta \delta$ and $\alpha \beta \delta \beta$ pattern to study the stable spacing regimes of the different patterns.
Eventually, we carry out a set of extended simulations starting from a random initial configuration of small solid-phase lamellae to explore the final pattern selection. 
The extended simulations are conducted by imposing periodic boundary conditions on the domain boundaries lying parallel to the growth direction.
Along the growth direction, the simulation box has a height of $\approx 3.6D_{max}/v$, where $D_{max} = max(D_{AA}^l, D_{BB}^l)$, and the moving window technique is imposed in the growth direction \cite{vondrous2014parallel} for simulating growth of the solid in an infinite reservoir of liquid maintained at the eutectic composition.

\section{Results}
\label{sec:results}
\subsection{Constrained simulations}
\label{sec:constrained}
\begin{figure}[htbp!]
  \centering
  \begin{subfigure}[b]{0.45\linewidth}
  \centering
    \includegraphics[width=0.95\linewidth]{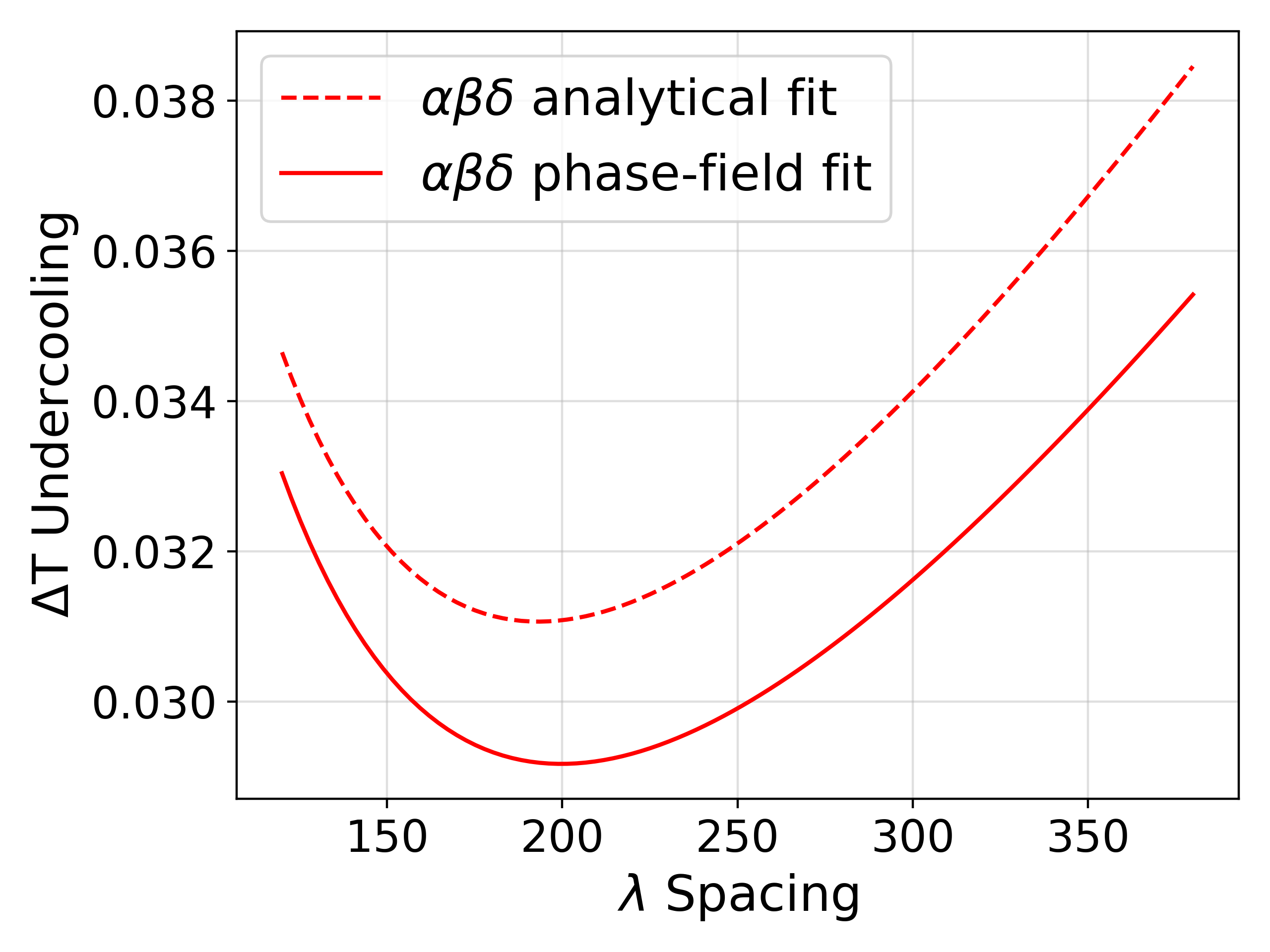}
    \caption{}
    \label{fig:pf_analy_abc}
  \end{subfigure}  
  \begin{subfigure}[b]{0.45\linewidth}
  \centering
    \includegraphics[width=0.95\linewidth]{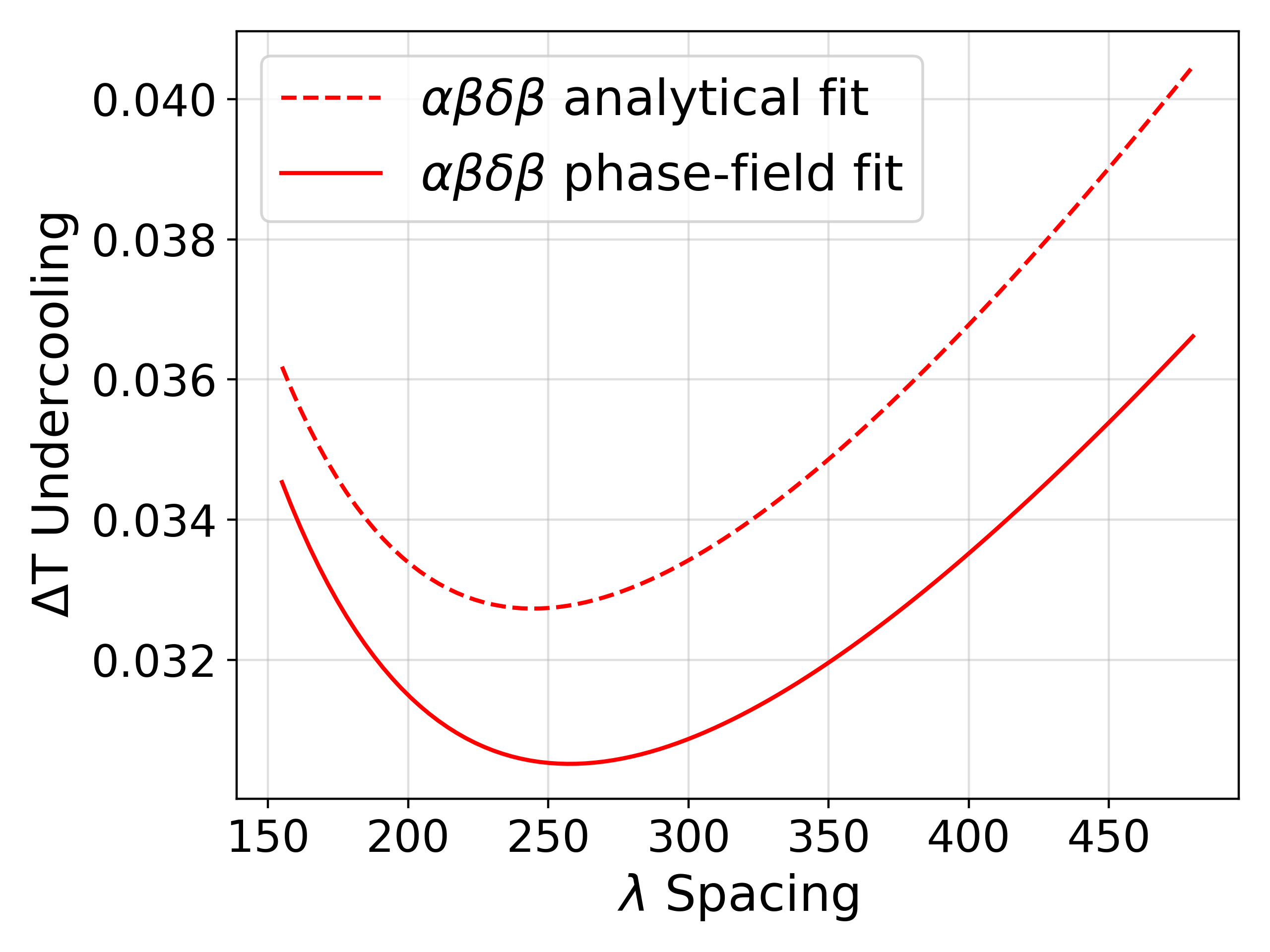}
    \caption{}
    \label{fig:pf_analy_abcb}
  \end{subfigure}
    \caption{Undercooling vs. spacing ($\Delta T - \lambda$) using the phase-field method and analytical solution for (a) $\alpha \beta \delta$ (b) $\alpha \beta \delta \beta$.}
  \label{fig:pf_analy}
\end{figure}

\begin{figure}[htbp!]
  \centering
  \begin{subfigure}[b]{0.3\linewidth}
  \centering
    \includegraphics[width=1\linewidth]{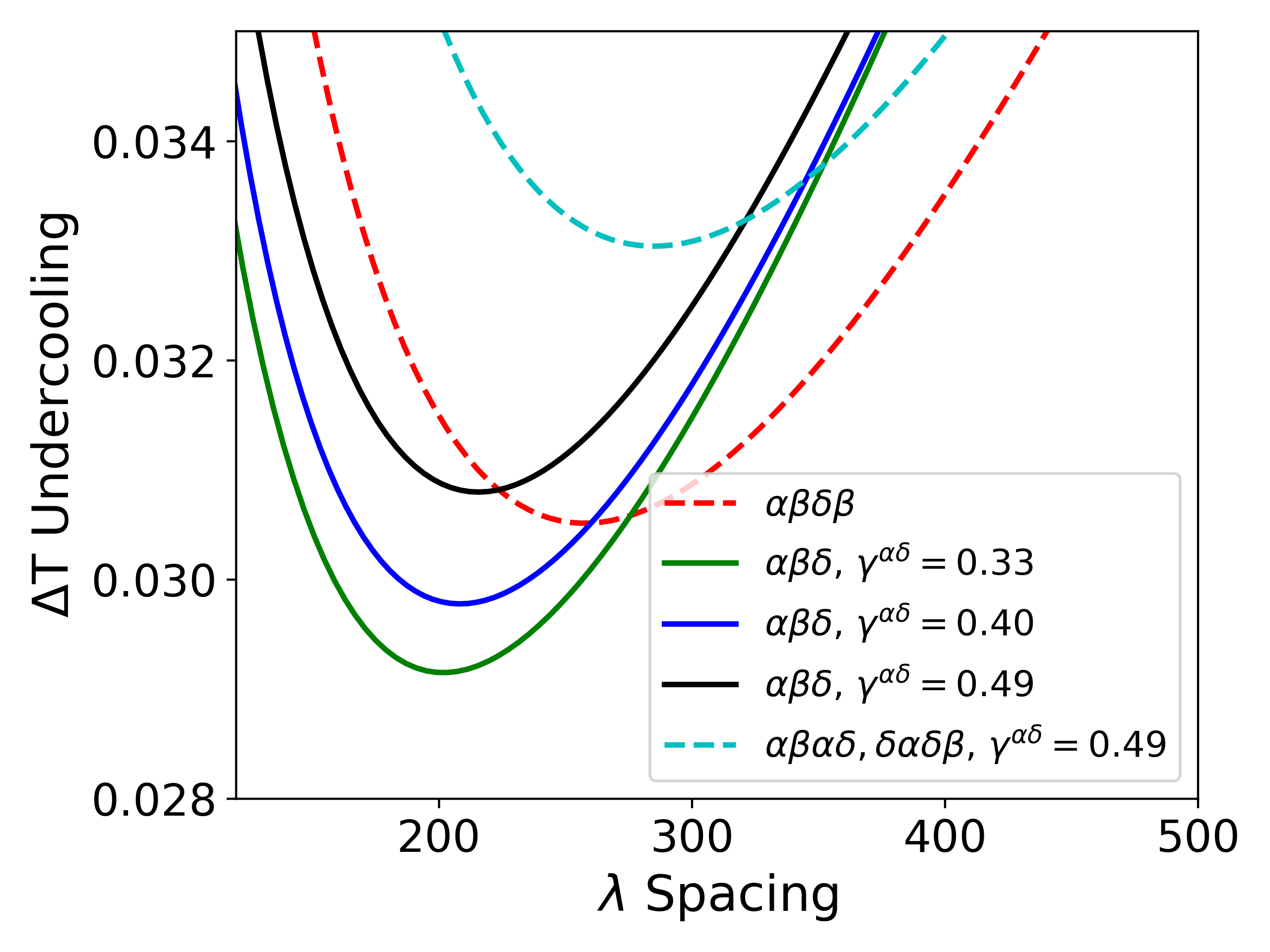}
    \caption{}
    \label{fig:diff_eq}
  \end{subfigure}  
  \begin{subfigure}[b]{0.3\linewidth}
  \centering
    \includegraphics[width=1\linewidth]{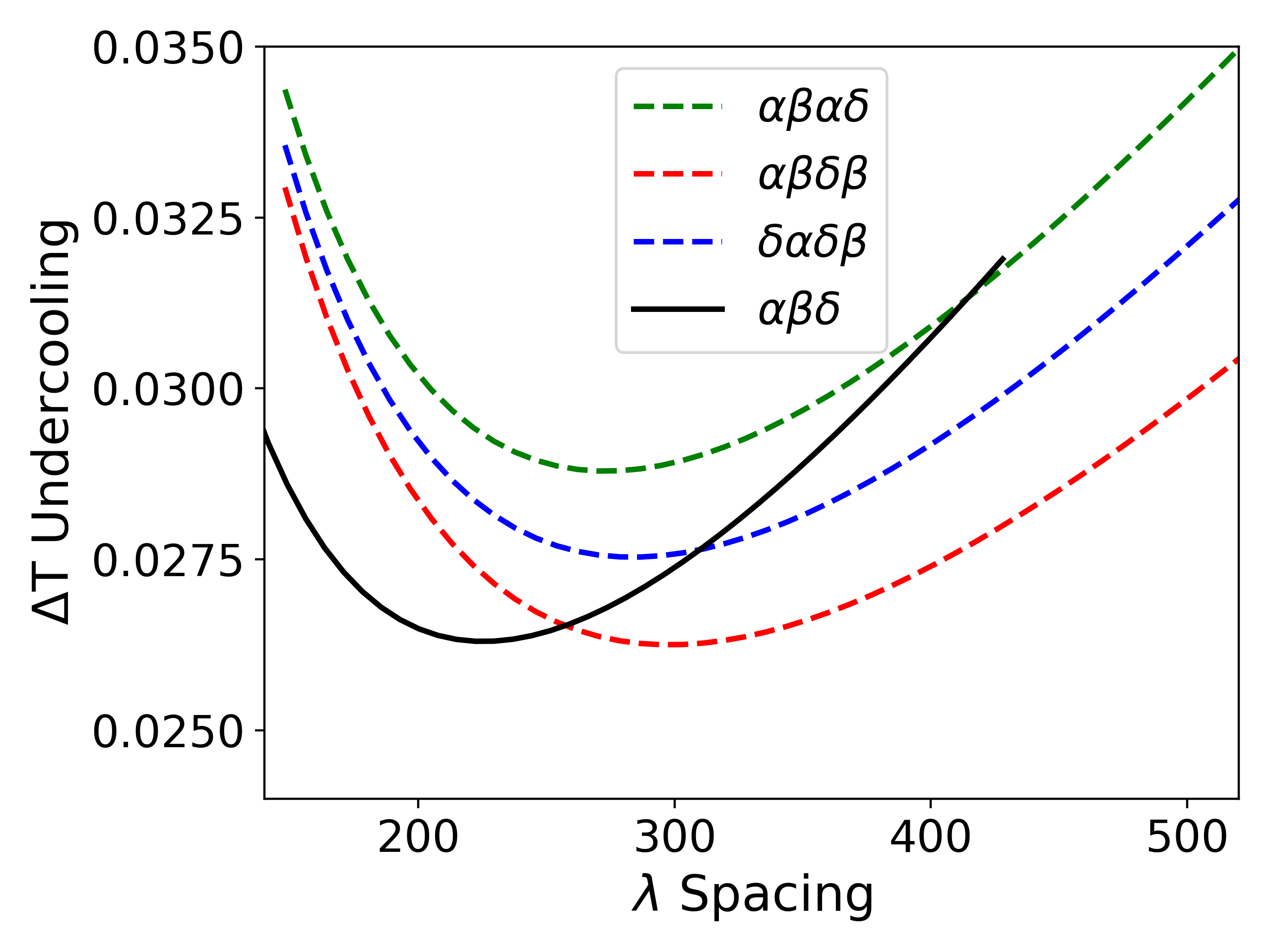}
    \caption{}
    \label{fig:diff_uneq_1p6}
  \end{subfigure}
   \begin{subfigure}[b]{0.3\linewidth}
   \centering
    \includegraphics[width=1\linewidth]{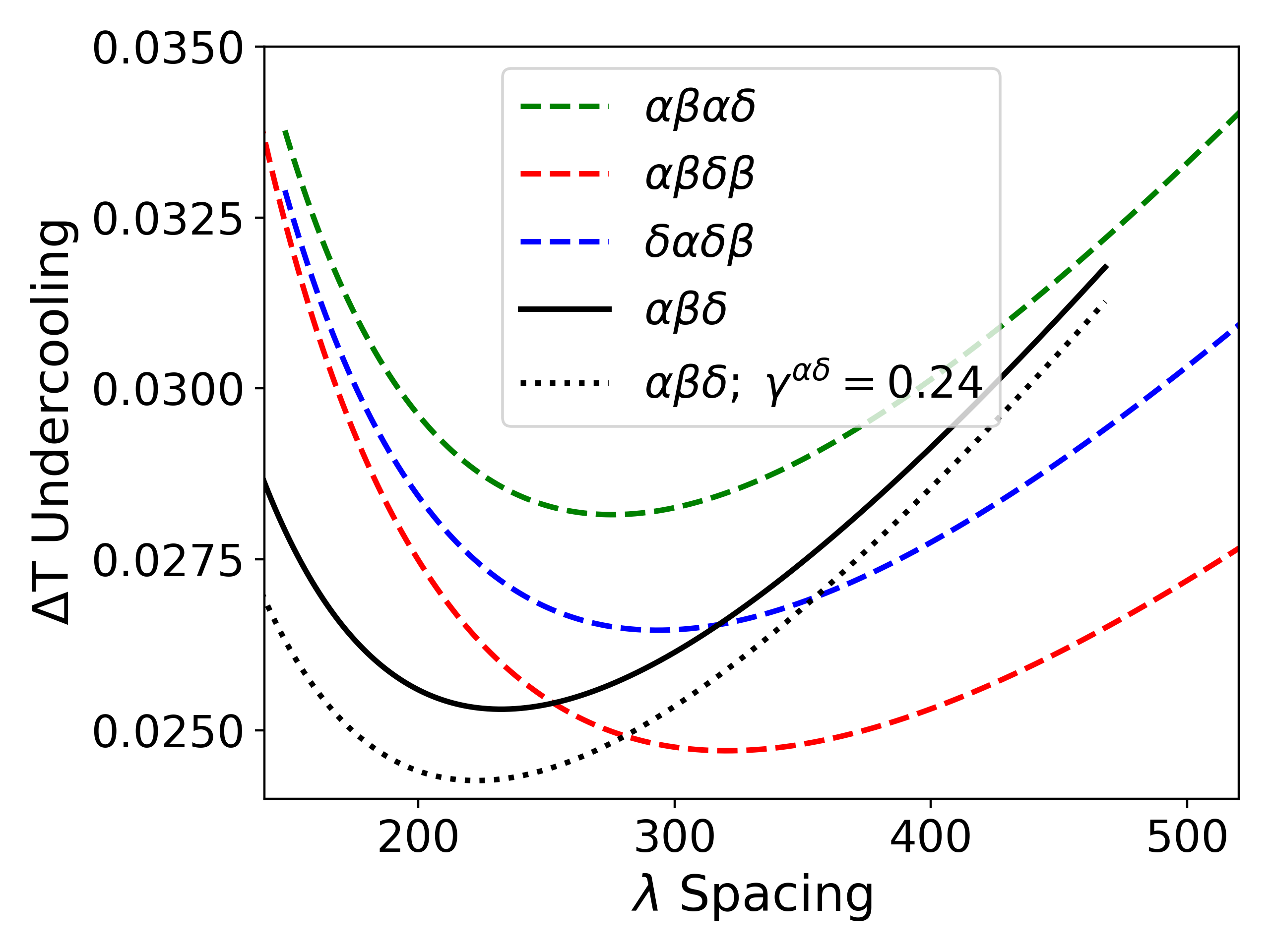}
    \caption{}
    \label{fig:diff_uneq}
  \end{subfigure}
  
      \begin{subfigure}[b]{0.48\linewidth}
   \centering
    \includegraphics[width=1\linewidth]{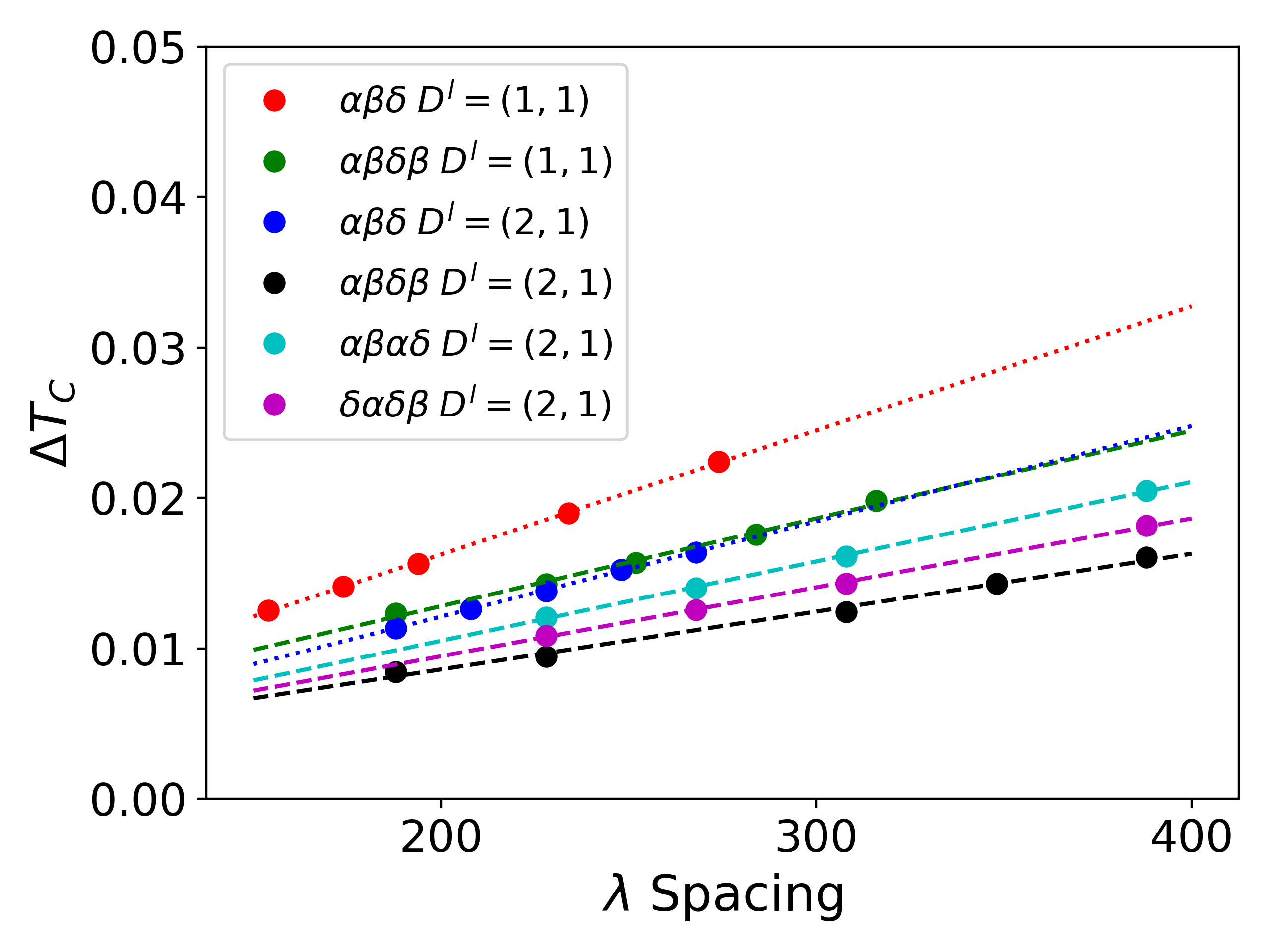}
    \caption{}
    \label{fig:undercooling_constitutional_alpha_beta_delta_avg}
  \end{subfigure}
    \begin{subfigure}[b]{0.48\linewidth}
   \centering
    \includegraphics[width=1\linewidth]{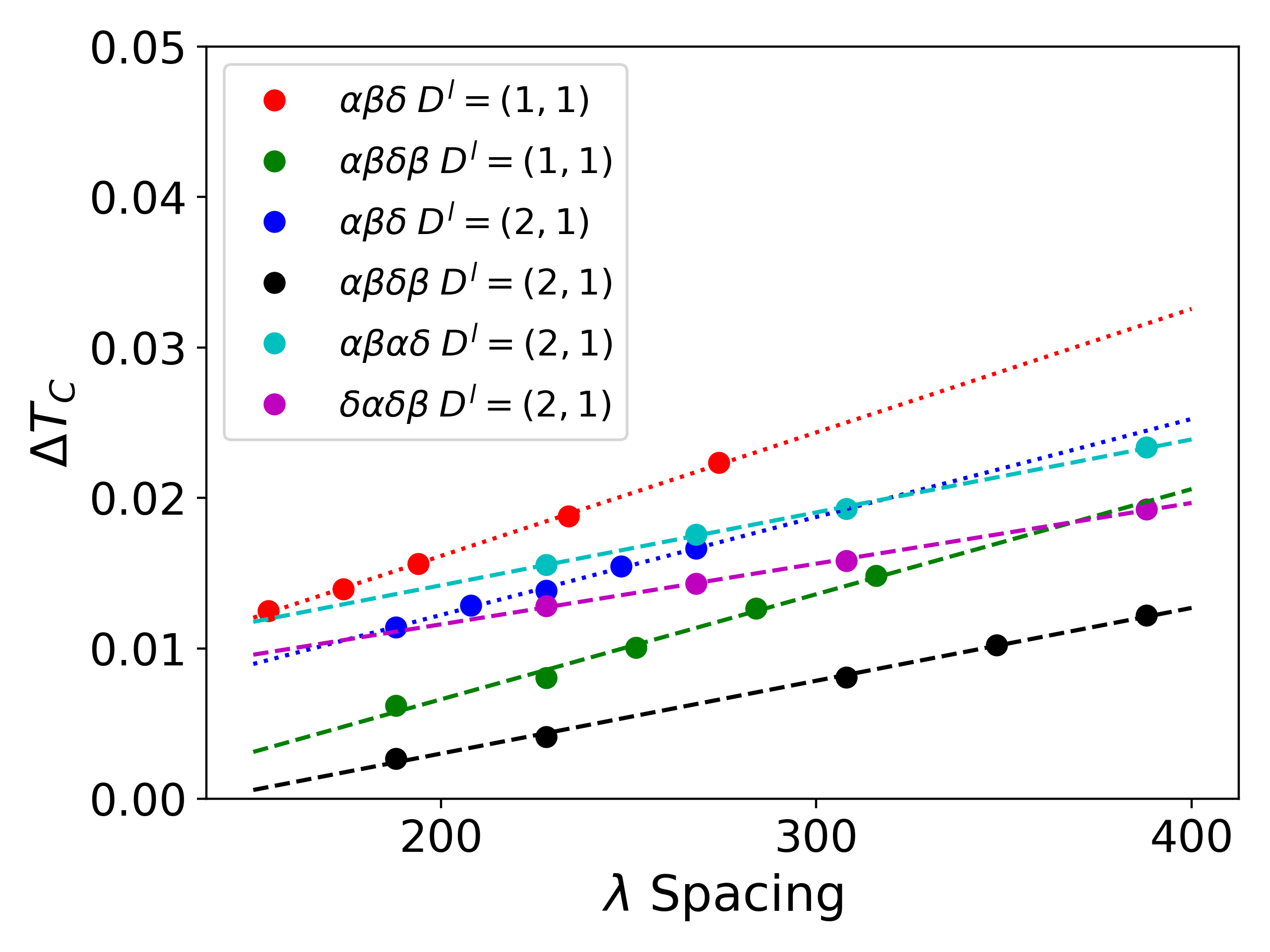}
    \caption{}
    \label{fig:undercooling_constitutional_beta}
  \end{subfigure}
    \caption{Undercooling vs. spacing ($\Delta T - \lambda$) plots from constrained simulations for (a) $D^l = (1, 1)$ with increasing interface energy of the $\alpha-\delta$ interface, from $0.333$ to $0.40$ and $0.49$. (b) $D^l = (1.6, 1)$ (c) $D^l = (2, 1)$. An increase in $\alpha-\delta$ interface energy or a relative lowering of the diffusivity of the B component leads to a lower minimum undercooling of the $\alpha \beta \delta \beta$ configuration compared to the $\alpha \beta \delta$. (d) Average of the constitutional undercooling of each phase extracted from phase-field simulations, dashed lines are linear fit. (e) Constitutional undercooling at the $\beta$-liquid interface extracted from phase-field simulations, dashed lines are linear fit.}
  \label{fig:spacing_plot}
\end{figure}

For the lamellar three-phase eutectics, the configurations having the lowest number of lamellae in a cycle are the $\alpha \beta \delta$ and $\alpha \beta \delta \beta$ (along with permutations).
We first investigate the solid-liquid interfacial undercooling vs. spacing ($\Delta T - \lambda$) variation for the $\alpha \beta \delta$ and $\alpha \beta \delta \beta$ lamellar arrangements of the three-phase eutectic patterns, and verify against the analytical solutions of the Jackson-Hunt type $\Delta T = K_1v\lambda + K_2/\lambda$, where $K_1, K_2$ depend upon the physical properties of the alloy system and the lamellar geometry as derived in \cite{choudhury2011theoretical}. 
The simulation domain is initialized with one $\alpha \beta \delta$ lamellar cycle with periodic boundary conditions imposed at the domain boundaries lying parallel to the growth direction. The $\alpha \beta \delta \beta$ pattern transforms to the $\alpha \beta \delta$ pattern for small spacings as observed in \cite{choudhury2011theoretical}, where the transformation, involves one of $\beta$ lamellae increasing in width while the other reduces and thus the same volume fraction of the phases is maintained. This transformation mechanism leads to elimination of one of the $\beta$ lamellae in the sequence, below a critical spacing which we will refer to as the short-wavelength instability spacing limit ($\lambda_{el}^{sw}$). Due to this, we initialize the simulation domain with just half the lamellar spacing of $\alpha \beta \delta \beta$ lamella with Neumann boundary conditions imposed at the domain boundaries lying parallel to the growth direction which are also the planes bisecting the $\alpha$ and $\delta$ lamellae. 
This constraint restricts the transformation of $\alpha \beta \delta \beta$ to $\alpha \beta \delta$ pattern, thus enabling us to map the entire undercooling vs. spacing variation of the $\alpha \beta \delta \beta$ morphology.
For equal diffusivities and interface energies, the patterns having four lamellae in a cycle, namely the $\alpha \beta \alpha \delta$, $\alpha \beta \delta \beta$ and $\delta \alpha \delta \beta$ patterns all have the same undercooling-spacing relationship due to the symmetry of the eutectic system chosen in our study, hence we will be referring to all these patterns as $\alpha \beta \delta \beta$.
The difference between the $\lambda_{min}$ derived from analytical predictions\cite{choudhury2011theoretical} and those derived from phase-field simulations for the $\alpha \beta \delta$ and $\alpha \beta \delta \beta$ patterns is $< 5\%$ as shown in Figure \ref{fig:pf_analy}.
The difference between the undercooling vs. spacing curves obtained from the analytical and phase-field simulations is attributed to the approximations in the analytical solution, and similar deviations between the two approaches have been discussed in References \cite{folch2005quantitative, choudhury2011theoretical, lahiri2017revisiting} for two-phase and three-phase eutectics.
The value of the minimum undercooling spacing obtained from the phase-field simulations, $\lambda_{min}^{\alpha \beta \delta} = 202$, and $\lambda_{min}^{\alpha \beta \delta \beta} = 258$ will be used to normalize the spacings in the rest of the paper.
As observed in Figure \ref{fig:diff_eq}, the $\Delta T_{min}$ for the $\alpha \beta \delta$ arrangement is lower than that for the $\alpha \beta \delta \beta$ arrangement for the case when all the interface energies are the same ($\gamma^{\alpha\beta} = \gamma^{\alpha\delta} = \gamma^{\beta\delta} = 0.333$). 
Next we impose asymmetries in the solid-solid interfacial energies and evaluate the undercooling vs. spacing variation for the $\alpha \beta \delta$ and $\alpha \beta \delta \beta$ patterns.
Upon increasing the $\alpha-\delta$ interface energy to $\gamma^{\alpha\delta} = 0.40$ and further to $\gamma^{\alpha\delta} = 0.49$, the $\alpha \beta \delta$ pattern goes from a lower minimum undercooling to a higher minimum undercooling compared to that of the $\alpha \beta \delta \beta$ pattern as depicted in Figure \ref{fig:diff_eq}. This can be understood on the basis of a lower curvature undercooling as one of the higher energy solid-solid interfaces $\alpha-\delta$ is absent from the $\alpha\beta\delta\beta$ pattern.
Similarly, an increase in the undercooling curve is observed for the $\alpha \beta \alpha \delta$ and $\delta \alpha \delta \beta$ patterns upon increasing the $\gamma^{\alpha\delta}$ to 0.49.
Another way we can add an asymmetry to our ideal symmetric eutectic system is by making the two solutal diffusivities of the liquid unequal, $D^l_{AA} > D^l_{BB}$, thus reducing the relative diffusivity of component B, which has the highest composition in the $\beta$ phase.
We observe a lower minimum undercooling for the pattern that has a higher number of the $\beta$ phase lamellae in a periodic cycle ($\alpha \beta \delta \beta$ pattern) compared to the other patterns with a lower $\beta$ phase lamellar frequency as shown in Figure  \ref{fig:diff_uneq} for a diffusivity ratio of the components $D^l_{AA}/D^l_{BB} = 2$.
The relative position of the undercooling vs. spacing plots can be explained based upon the changes in the average constitutional undercooling extracted from the phase-field simulations shown in Figure  \ref{fig:undercooling_constitutional_alpha_beta_delta_avg}, which reveals that the average constitutional undercooling of $\alpha\beta\delta\beta$ configuration with the highest frequency of the $\beta$ lamellae that is also rich in the slowest diffusing component has the lowest constitutional undercooling. The constitutional undercooling is lowest at the $\beta$ liquid interface as highlighted in Figure \ref{fig:undercooling_constitutional_beta} for the $\alpha\beta\delta\beta$ pattern. Further, by increasing the diffusivity of one of the species while keeping the other diffusivity constant leads to an increase in the effective diffusion length in the system. This results in a reduction in the slopes of the variation of the constitutional undercooling with spacing for all patterns which leads to the shift in the minimum undercooling spacing to larger values with the maximum shift occurring for the $\alpha\beta\delta\beta$ configuration.
\begin{figure}[htbp!]
  \centering
    \includegraphics[width=0.5\linewidth]{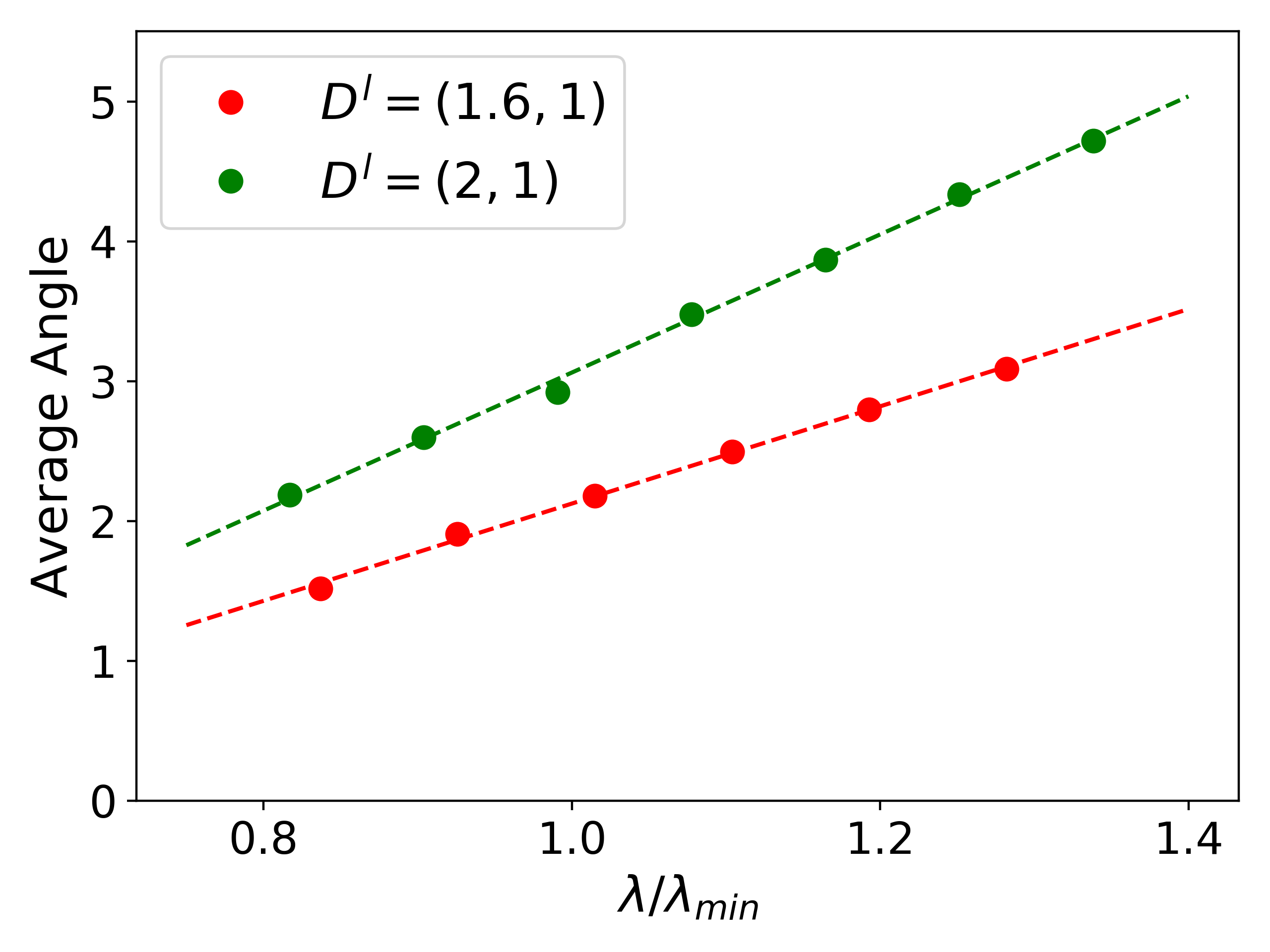}
    \caption{Average tilt angle (in degrees) of solid-solid interface with respect to the temperature gradient direction for $\alpha \beta \delta$ pattern with $D^l = (1.6, 1)$ and $D^l = (2, 1)$. Spacing scaled with the $\lambda_{min}$ corresponding to the respective diffusivity values for the $\alpha \beta \delta$ pattern.}
  \label{fig:D2_angle_ABC}
\end{figure}
With an increase in diffusivities, the minimum undercooling spacing for $D^l = (2, 1)$   ($\lambda_{min,D_2}^{\alpha \beta \delta} = 233$, and $\lambda_{min,D_2}^{\alpha \beta \delta \beta} = 320$) has increased compared to the case with equal diffusivities.
Further, the steady state growth morphology for the $\alpha\beta\delta$ pattern bears a tilt with respect to the direction of the thermal gradients which is possibly due to the absence of a mirror-symmetry in the $\alpha \beta \delta$ that has also been reported previously \cite{lahiri2017revisiting}. The variation of the average tilt of the solid-solid interfaces with spacing is highlighted in Figure \ref{fig:D2_angle_ABC} for $D^l = (1.6, 1)$ and $D^l = (2, 1)$. 
The average tilt angle increases linearly for spacings around $\lambda_{min}$ for both the diffusivity ratios, with a higher tilt angle for $D^l = (2, 1)$ compared to $D^l = (1.6, 1)$.
For the $\alpha \beta \delta \beta$ pattern no tilt in the solid-solid interface is observed, due to the presence of a mirror symmetry bisecting the $\alpha$ and $\delta$ lamellae.

\begin{figure}[htbp!]
  \centering
  \begin{subfigure}[b]{0.24\linewidth}
  \centering
    \includegraphics[height=7cm]{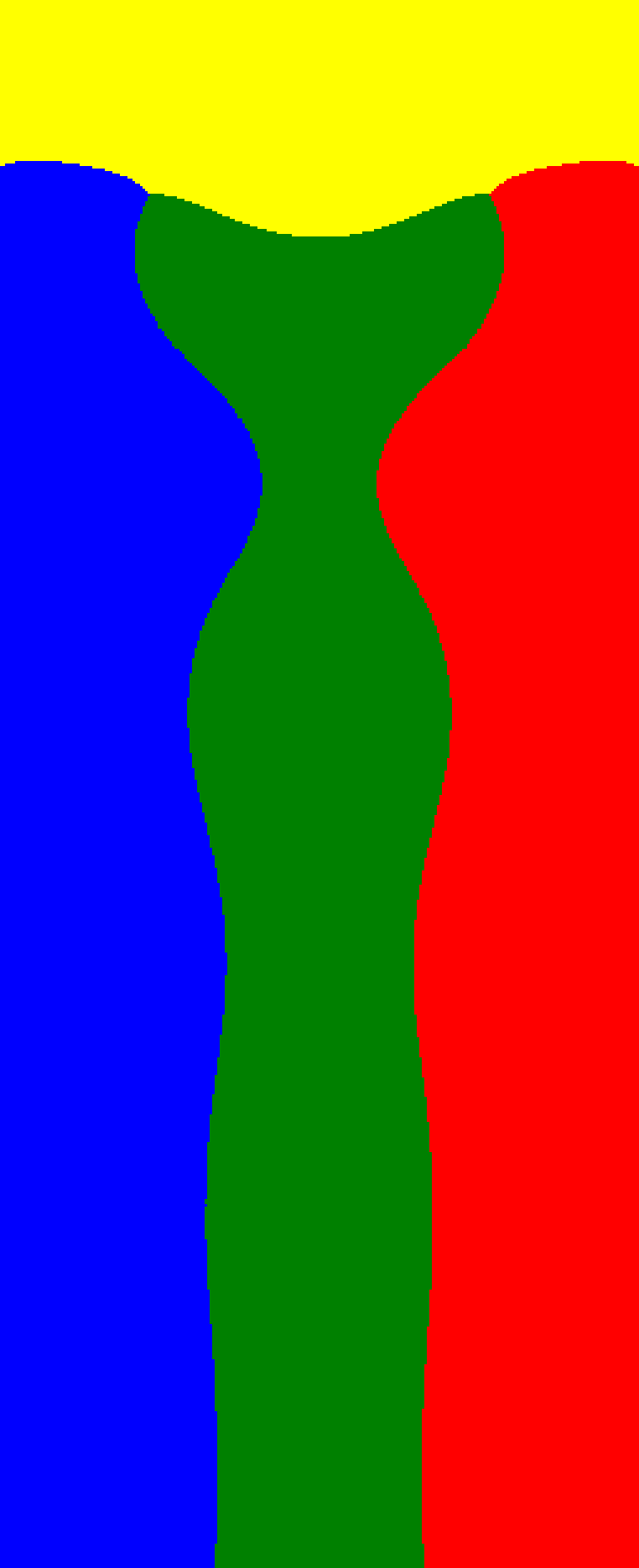}
    \caption{}
    \label{fig:oscillation_D1_ABC}
  \end{subfigure}
  \begin{subfigure}[b]{0.24\linewidth}
  \centering
    \includegraphics[height=7cm]{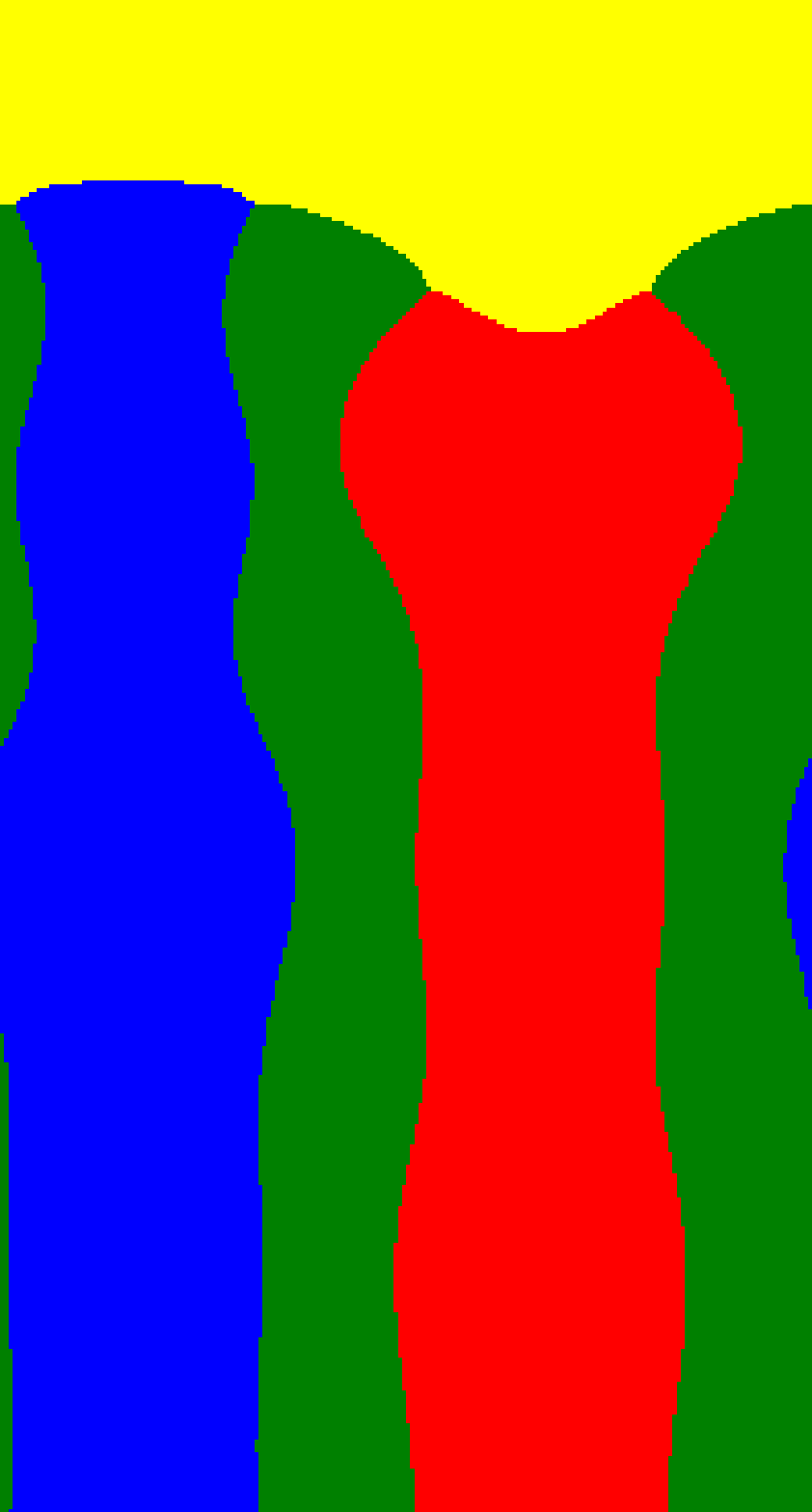}
    \caption{}
    \label{fig:oscillation_D1_ABAC}
  \end{subfigure}
  \begin{subfigure}[b]{0.24\linewidth}
  \centering
    \includegraphics[height=7cm]{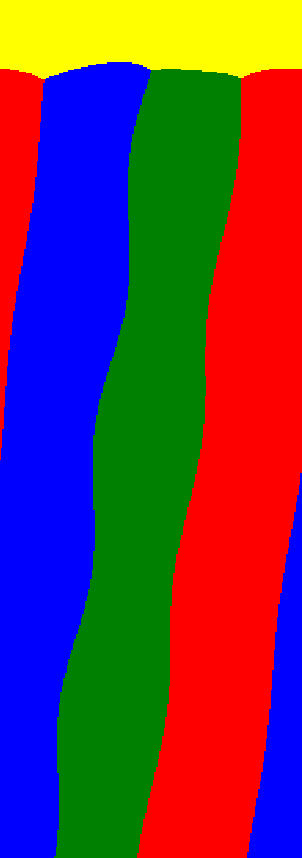}
    \caption{}
    \label{fig:oscillation_D2_ABC}
  \end{subfigure}
    \begin{subfigure}[b]{0.24\linewidth}
  \centering
    \includegraphics[height=7cm]{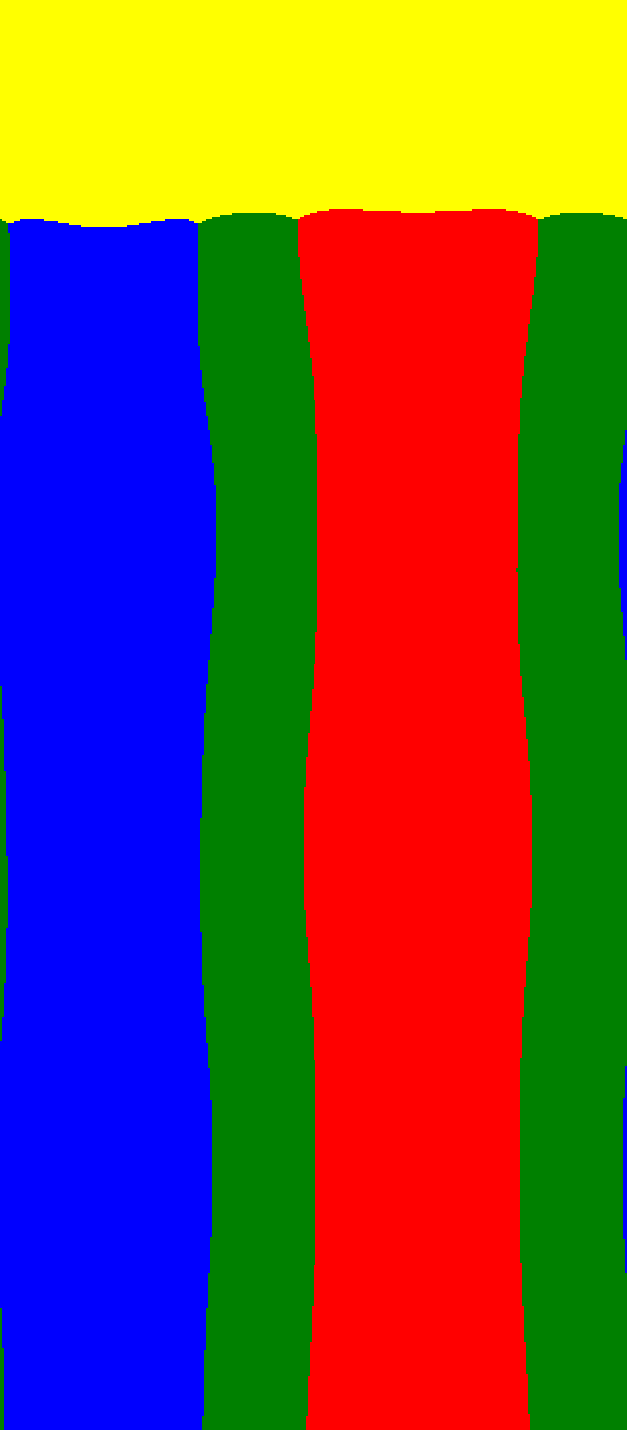}
    \caption{}
    \label{fig:oscillation_D2_BABC}
  \end{subfigure}
    \caption{Oscillatory modes for simulations at larger spacings for different patterns and diffusivities (a) $1-\lambda-\text{O}$ mode for $D^l = (1, 1)$, $\alpha \beta \delta$; (b) mixed mode ($1-\lambda-\text{O}$ and $2-\lambda-\text{O}$) for $D^l = (1, 1)$, $\alpha \beta \delta \beta$; (c) $1-\lambda-\text{O}$ mode with tilt for $D^l = (2, 1)$, $\alpha \beta \delta$; (d) $2-\lambda-\text{O}$ mode for $D^l = (2, 1)$, $\alpha \beta \delta \beta$.}
  \label{fig:oscillation}
\end{figure}

For much higher spacings, we observe different oscillatory modes depending upon the symmetry of the pattern (Figure \ref{fig:oscillation}) which is explored in detail in \cite{choudhury2011theoretical}.
The $\alpha \beta \delta$ pattern shows $1-\lambda-\text{O}$ mode oscillations for equal diffusivities, and $1-\lambda-\text{O}$ mode with tilt for unequal diffusivities, (Figure \ref{fig:oscillation_D2_ABC} for $D^l = (2, 1)$).
The $\alpha \beta \delta \beta$ shows a mixed mode ($1-\lambda-\text{O}$ and $2-\lambda-\text{O}$) for equal diffusivities, and a $2-\lambda-\text{O}$ mode where the width of the $\beta$ phase lamellae remains the same while oscillating for unequal diffusivities. Since, the 
present paper is devoted to the study of pattern competition and selection we will defer the determination of the complete stability diagrams and mapping of the oscillatory instabilities to a later study. 

Thus we have a combination of parameters where the minimum undercooling of the $\alpha \beta \delta \beta$ pattern and $\alpha \beta \delta$ pattern bear the following relationship, $\Delta T_{min}^{\alpha \beta \delta} < \Delta T_{min}^{\alpha \beta \delta \beta}$ for $D^l = (1, 1), \gamma^{\alpha \delta} = 0.333$ and $D^l = (2, 1), \gamma^{\alpha \delta} = 0.24$; $\Delta T_{min}^{\alpha \beta \delta} \approx \Delta T_{min}^{\alpha \beta \delta \beta}$ for $D^l = (1.6, 1), \gamma^{\alpha \delta}=0.333$ and $\Delta T_{min}^{\alpha \beta \delta} > \Delta T_{min}^{\alpha \beta \delta \beta}$ for $D^l = (1, 1), \gamma^{\alpha \delta} = 0.49$ and $D^l = (2, 1), \gamma^{\alpha \delta} = 0.333$. 
So far, using constrained simulations we have obtained a first insight on the undercooling vs. spacing relationships for the different lamellar patterns under varying solid-solid interfacial energies and liquid component diffusivities.
The subsequent results will focus on the pattern stability, pattern transition and selection for the mentioned five parameter sets under extended simulations. 

\subsection{Extended simulations}
\label{sec:extended}
In this section we investigate three aspects about the configurations, first is the stability of the simplest configurations to long-wavelength perturbations, followed by studies on pattern competition and pattern selection. For this, we perform extended simulations with a large number of lamellae, where given the higher degrees of freedom for lamella adjustment, dynamic properties like the pattern stability to long-wavelength perturbations, competition and transformation between different patterns are captured.

\subsubsection{Pattern stability}
\label{sec:stability}
\begin{figure}[htbp!]
  \centering
    \includegraphics[width=0.5\linewidth]{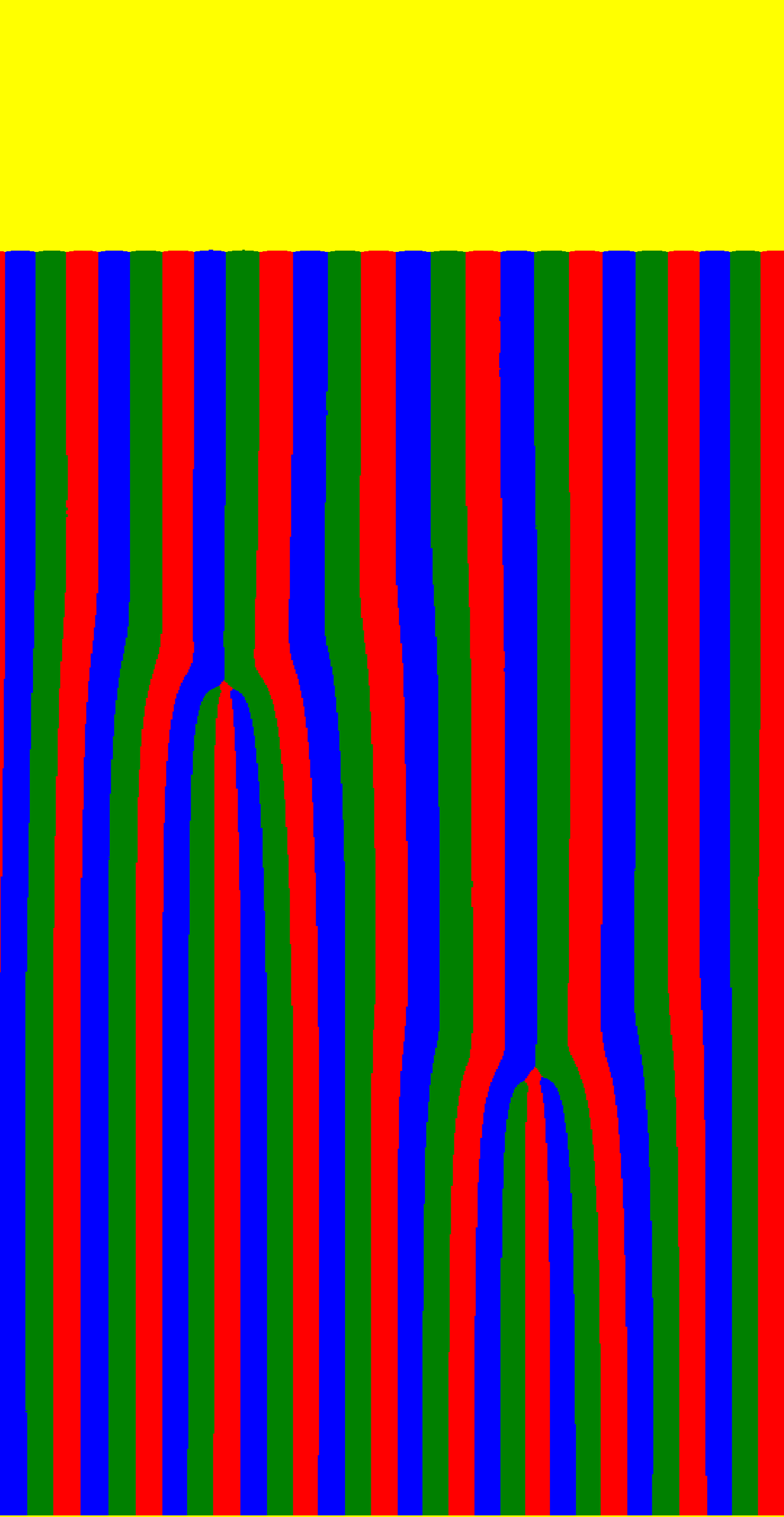}
    \caption{$\alpha \beta \delta$ morphology of average spacing $\lambda = 162 = 0.8\lambda^{\alpha \beta \delta}_{min}$ and a small random noise imparted to the initial spacing. Elimination of 2 periods of $\alpha \beta \delta$ pattern is observed. Image is scaled 2X in the width. In all the microstructural images in this paper, the red, green, blue and yellow colors represent the $\alpha$, $\beta$, $\delta$ and liquid phases respectively.}
    \label{fig:D1_c7_my_544_big}
   \end{figure}

First, we study the stability of the patterns to long-wavelength perturbations in the lamellar spacing. 
For this, we initialize the morphology with a large number of periods ($N = 10\; \text{or}\; 20$) of the $\alpha \beta \delta$ and $\alpha \beta \delta \beta$ patterns with different domain sizes corresponding to different average spacings.
To this configuration of lamellae of uniform width, a small random deviation is imparted to the width of each lamella.
Figure \ref{fig:D1_c7_my_544_big} shows the evolution of the $\alpha \beta \delta$ morphology starting with 10 periods with an average spacing of $\lambda = 162 = 0.8\lambda^{\alpha \beta \delta}_{min}$, which is just below the threshold for lamella elimination ($\lambda^{\alpha \beta \delta}_{el} = 0.83\lambda^{\alpha \beta \delta}_{min}$). 
Elimination of 2 $\alpha \beta \delta$ periods is observed resulting in the steady state configuration having 8 $\alpha \beta \delta$ periods and an increased average $\alpha \beta \delta$ spacing of $\lambda = 202 = \lambda^{\alpha \beta \delta}_{min}$.
Thus the $\alpha \beta \delta$ pattern undergoes lamellar elimination due to a long-wavelength perturbation in spacing, whereas the $\alpha \beta \delta \beta$ pattern undergoes transition to an $\alpha \beta \delta$ pattern for spacings less than $1.09\lambda^{\alpha \beta \delta \beta}_{min}$ by elimination of alternate $\beta$ phase lamella (Figure \ref{fig:D1_config_2_t_18L}). This particular transition occurs even for a single period of $\alpha\beta\delta\beta$, therefore, the lower bound of spacings for the $\alpha\beta\delta\beta$ is limited by this short-wavelength instability \cite{choudhury2011theoretical} and not the true lamella elimination instability limit due to an Eckhaus type instability. 
   
\begin{figure}[htbp!]
  \centering
  \begin{subfigure}[b]{0.32\linewidth}
  \centering
    \includegraphics[width=0.95\linewidth]{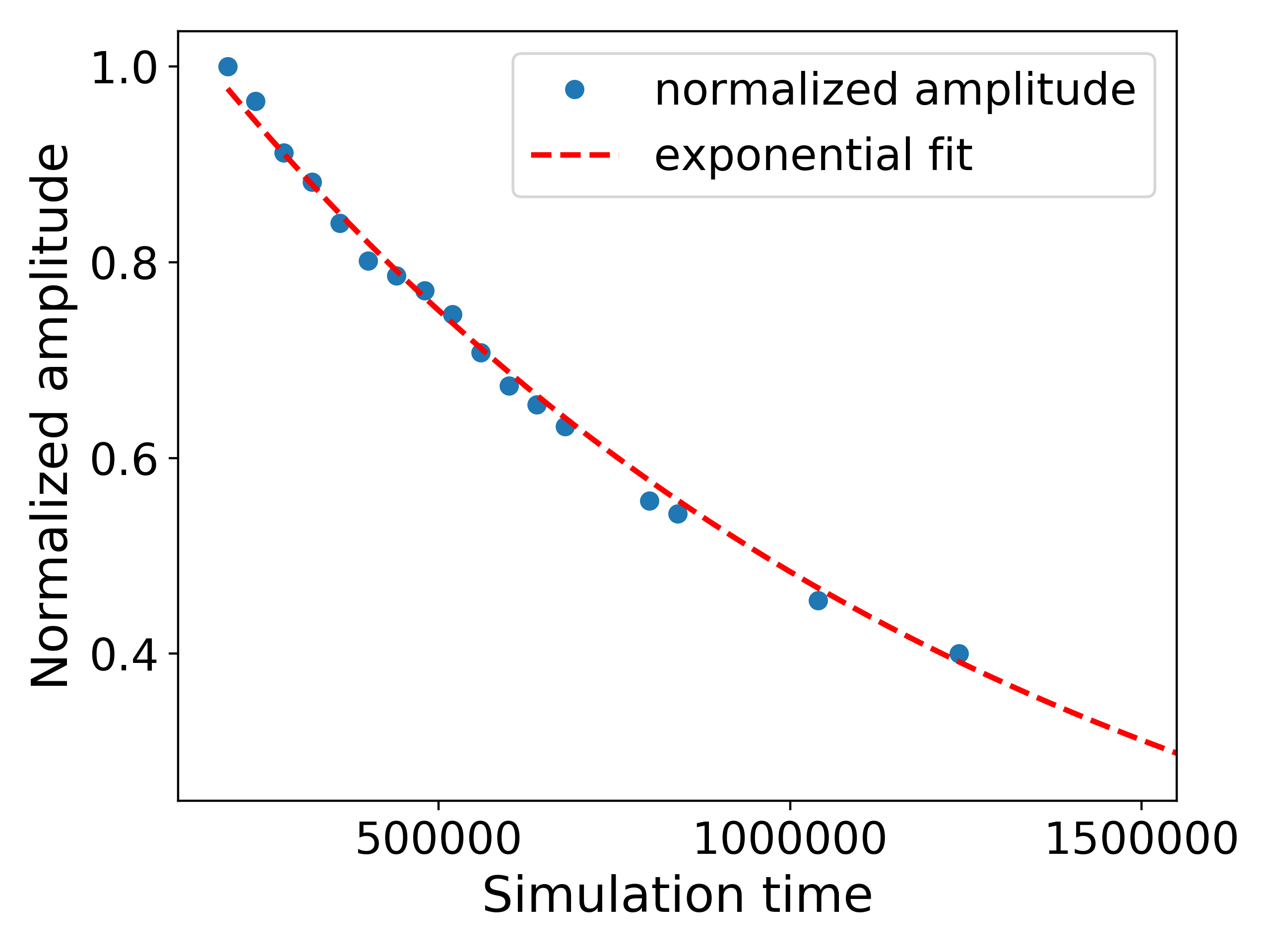}
    \caption{}
    \label{fig:pd_exp}
  \end{subfigure}  
  \begin{subfigure}[b]{0.32\linewidth}
  \centering
    \includegraphics[width=0.95\linewidth]{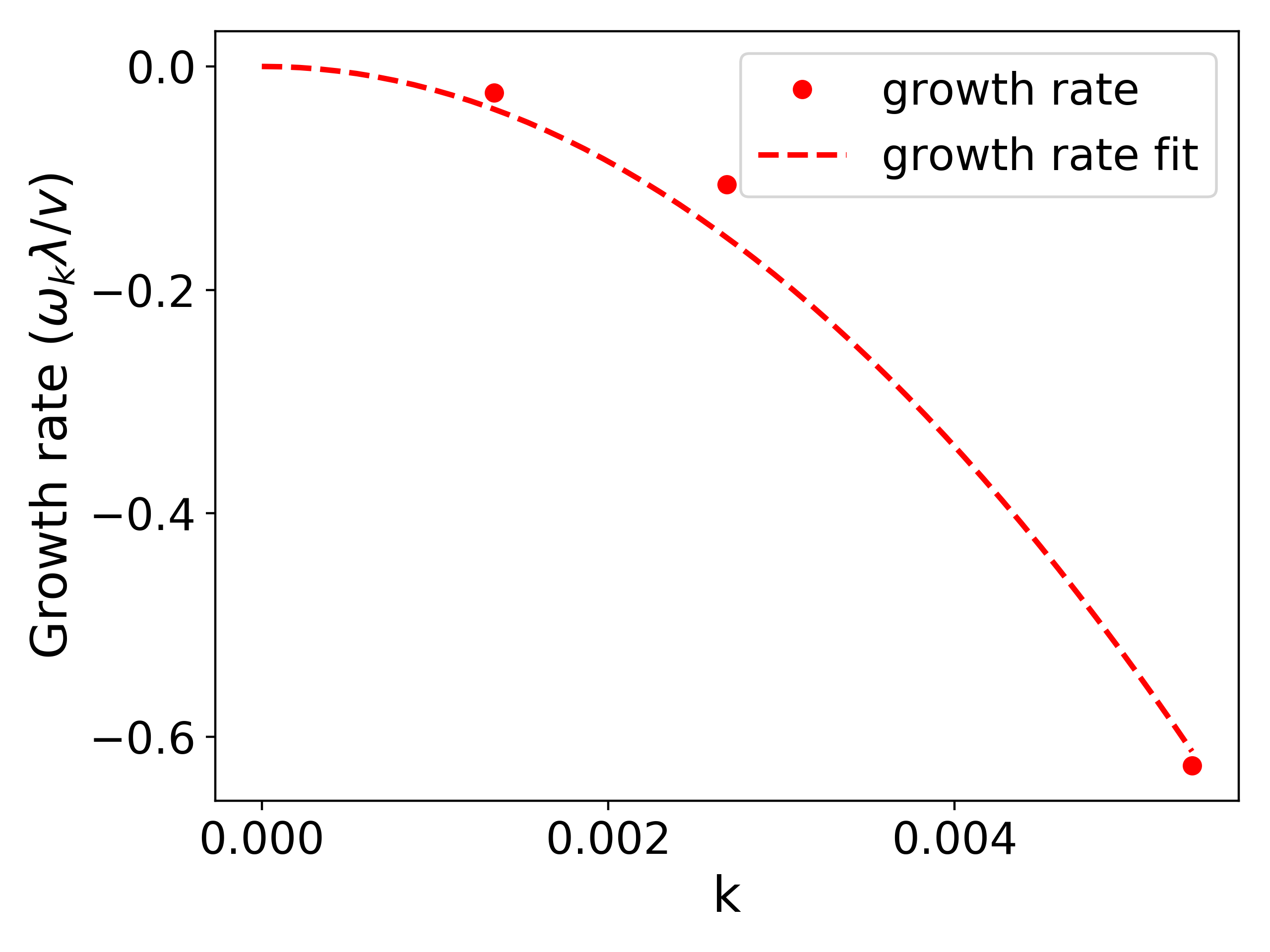}
    \caption{}
    \label{fig:pd_parabola}
  \end{subfigure}
  \begin{subfigure}[b]{0.32\linewidth}
  \centering
    \includegraphics[width=0.95\linewidth]{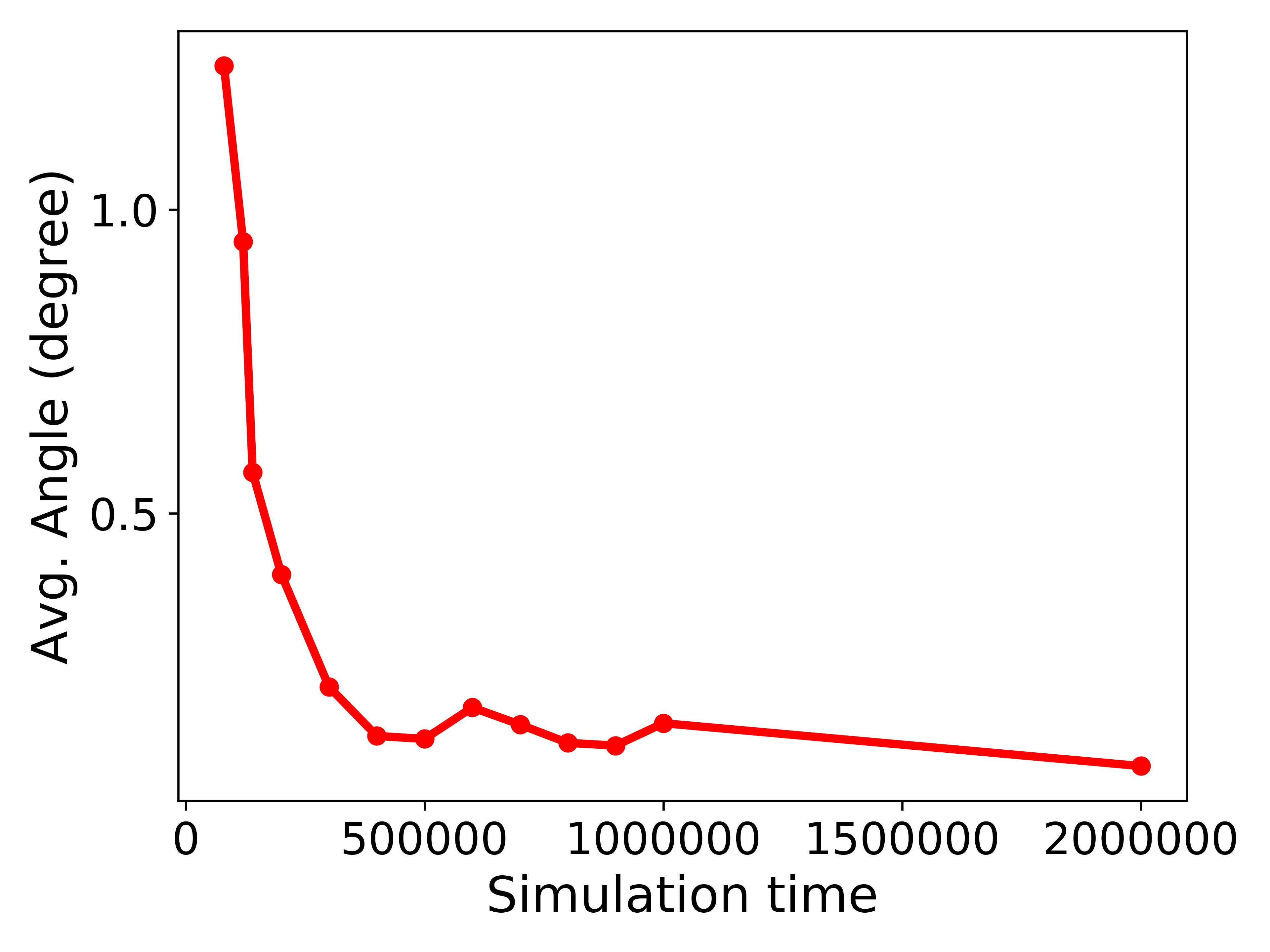}
    \caption{}
    \label{fig:pd_angle}
  \end{subfigure}
    \caption{(a) Decaying exponential fit to Fourier amplitude with time to determine $\omega_k$. (b) Parabola fit to growth rate ($\omega_{k} = -D_{\lambda} k^2$) to determine the phase-diffusion coefficient $D_\lambda$ for $\lambda = 1.16\lambda_{min}^{\alpha\beta\delta}$. (c) Average over all triple-points of the angle between the triple-point velocity vector and the local solid-liquid envelope normal. In (c), the line through the points is drawn as a guide to the eye.}
  \label{fig:pd_plots}
\end{figure}

For larger spacings beyond $\lambda_{el}$, we study the kinetics of spacing homogenization, and calculate the phase-diffusion coefficient.
The phase-diffusion coefficient indicates how fast the system stabilizes itself against a perturbation of the lamellar spacing. 
The spacings for each of the $\alpha \beta \delta$ and the $\alpha \beta \delta \beta$ patterns is calculated by tracking the $\alpha-\beta-l$ triple-point coordinates, denoted as $(x_n, y_n)$ for the n'th period, which is at the intersection of $\phi_{\alpha} -\phi_l = 0$ and $\phi_{\beta}- \phi_l = 0$ contours. 
From the $\alpha-\beta-l$ triple points we calculate the spacing of each period as $\lambda_{n} = x_{n} - x_{n-1}$, and find the deviation of the spacing ($\delta \lambda_{n}$) from the average spacing ($\lambda$), computed as $\delta \lambda_{n} = \lambda_{n} - \lambda$.
We obtain the discrete Fourier transform as \cite{akamatsu2004overstability}
\begin{flalign}
Y(k, t) = \frac{1}{N}\sum_{n=0}^{N-1}\delta \lambda_n exp\left(ink\lambda\right),
\label{eqn:fourier}
\end{flalign}
where k is the wave-vector. 
We fit the Fourier amplitude ($Y(k, t)$) corresponding to the small wave-vectors $k$ (or large wavelengths $L = 2\pi/k$) with an exponential in time ($Y(k, t) = exp(\omega_{k} t)$) to compute the growth rate ($\omega_k$) as shown in Figure \ref{fig:pd_exp}.
The phase-diffusion coefficient $D_{\lambda}$ for  average spacing $\lambda$ is obtained from the second order coefficient of parabolic fit of $\omega_{k} = -D_{\lambda} k^2$ for small $k$ in Figure \ref{fig:pd_parabola}.
The calculated phase-diffusion coefficient $D_{\lambda}$ is larger than the one obtained from the analytical solution in \cite{langer1980eutectic} given as 
\begin{equation}
D_{\perp} = \frac{K_{1}\lambda v^2}{G_T}\big(1-\frac{1}{{\Lambda}^2}\big),
\end{equation}
where $\Lambda = \lambda/\lambda_{min}$, $K_1$ is the coefficient of fit to the Jackson-Hunt equation of the form $\Delta T = K_1v\lambda + K_2/\lambda$ and the difference ($D_{\lambda} - D_{\perp}$) is attributed to the component of triple-point motion parallel to the solid-liquid envelope ($D_{||} = D_{\lambda} - D_{\perp}$) \cite{plapp2002eutectic}.
In order to investigate the motion of the triple junctions, we find the angle between the triple-point velocity vector and the local solid-liquid envelope normal for each triple-point. 
The average of the absolute of this angle for all the triple-points with time is plotted in Figure \ref{fig:pd_angle}.
Although the deviation is small $<\ang{1}$ as also observed in \cite{plapp2002eutectic, akamatsu2004overstability}, this leads to a  triple-junction velocity component parallel to the local solid-liquid envelope, which is responsible for the stability of the $\alpha \beta \delta$ pattern for spacings as low as $\lambda_{el} = 0.83\lambda_{min}^{\alpha \beta \delta}$.
Using the empirical equation $D_{||} = Bv\lambda \Lambda$ from \cite{akamatsu2002pattern, akamatsu2004overstability}, we get the value of the constant coefficient $B^{\alpha \beta \delta} \approx 0.05$ and $B^{\alpha \beta \delta \beta} \approx 0.11$. While we have investigated the pattern stability to long-wavelength perturbations for the case of equal diffusivities, for the case of unequal diffusivities we do not expect a change in the behavior of the temporal evolution of the spacings in the $\alpha\beta\delta$ pattern, with only a change in the growth exponents brought about by the modification to the undercooling vs. spacing relationships. However, for the $\alpha\beta\delta\beta$ pattern in addition to the change in phase diffusivities, a difference in the temporal evolution is expected for smaller spacings due to the change in the relative positions of the short-wavelength and Eckhaus instability limits brought about by the changes in the diffusivity contrast (for example, the short-wavelength instability limit reduces from $\lambda_{el}^{sw}=1.09\lambda_{min}^{\alpha\beta\delta\beta}$ for $D^l=(1,1)$ to $\lambda_{el}^{sw}=0.82\lambda_{min,D_2}^{\alpha\beta\delta\beta}$ for $D^l=(2,1)$).   
For parameters where the lamella elimination instability limit is larger than the short-wavelength instability limit, we expect to see an Eckhaus type instability also for the $\alpha\beta\delta\beta$ pattern. For such cases, there will be spacings between the limits corresponding to the short-wavelength instability and the lamella elimination instability where the Eckhaus instability will lead to an increase in spacings (without the coordinated elimination of $\beta$ lamellae as in the short-wavelength instability) and thereby the transition to the simpler $\alpha\beta\delta$ patterns will not occur.

\subsubsection{Pattern competition}
\label{sec:competition}
\begin{figure}[htbp!]
  \centering
  \begin{subfigure}[b]{0.9\linewidth}
  \centering
    \includegraphics[width=0.5\linewidth]{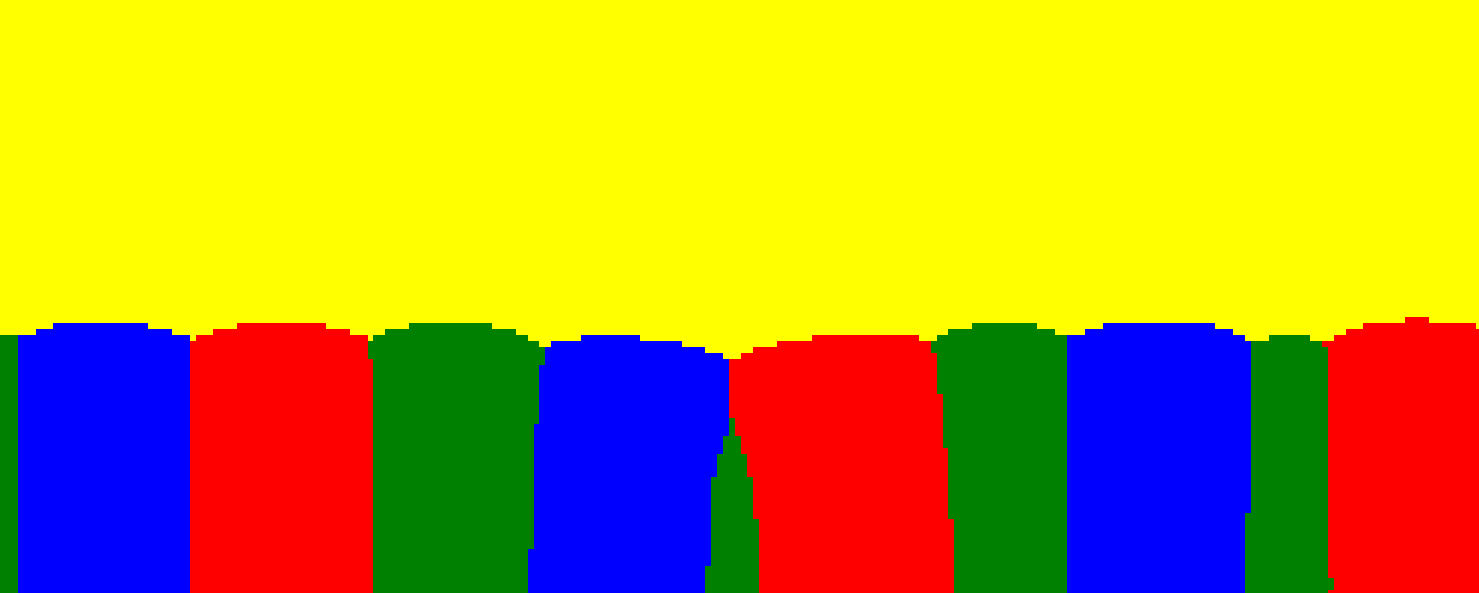}
    \caption{}
    \label{fig:D1_config_2_t_18L}
  \end{subfigure}
  
   \begin{subfigure}[b]{0.9\linewidth}
   \centering
    \includegraphics[width=0.9\linewidth]{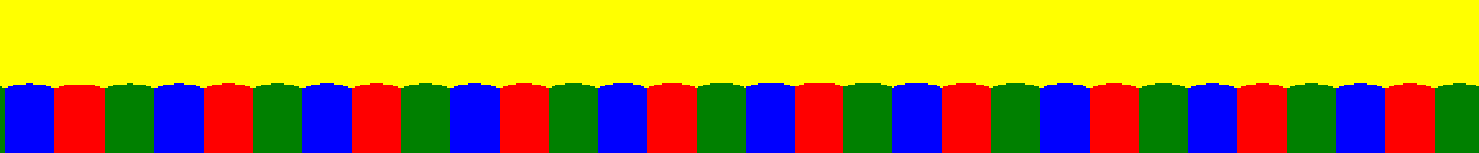}
    \caption{}
    \label{fig:D1_config_2_t_60L}
  \end{subfigure}
      \caption{(a) Zoomed in section showing the $\alpha \beta \delta \beta$ pattern transforming to $\alpha \beta \delta$
      pattern by elimination of alternate $\beta$ phase lamella for an initial spacing of $\lambda = 270$ for both the patterns with the initial configuration $[\alpha\beta\delta\beta]_5[\alpha\beta\delta]_5$. (b) Final steady state with the configuration $[\alpha \beta \delta]_{10}$ with $\lambda = 270 = 1.33\lambda_{min}^{\alpha \beta \delta}$.}
  \label{fig:D1_config_2}
\end{figure}

\begin{figure}[htbp!]
  \centering
  \begin{subfigure}[b]{0.9\linewidth}
  \centering
    \includegraphics[width=0.9\linewidth]{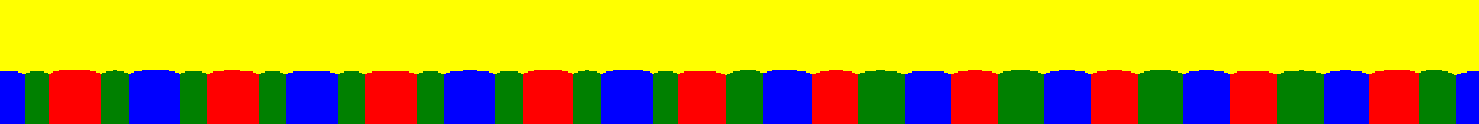}
    \caption{}
    \label{fig:D1_config_2_1204_no_transform}
  \end{subfigure}
  
    \begin{subfigure}[b]{0.9\linewidth}
    \centering
    \includegraphics[width=0.9\linewidth]{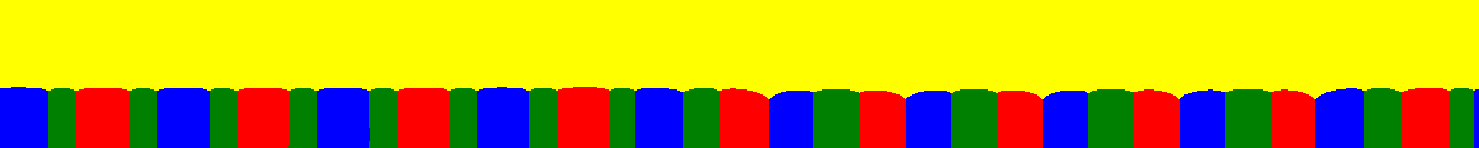}
    \caption{}
    \label{fig:D1_config_2_alpha_0p56_1204_no_transform}
  \end{subfigure}
      \caption{Starting with an initial spacing of $\lambda = 360$ for both the $\alpha \beta \delta$ and $\alpha \beta \delta \beta$ patterns with the configuration $ [\alpha\beta\delta\beta]_5[\alpha\beta\delta]_5$. (a) Both the patterns $\alpha \beta \delta$ and $\alpha \beta \delta \beta$ coexist at the same interfacial undercooling. (b) For a higher interfacial energy of $\gamma^{\alpha \delta} = 0.49$, still both the patterns coexist.}
  \label{fig:D1_config_2_1204_and_schematic}
\end{figure}

In order to study competition between the $\alpha \beta \delta$ and $\alpha \beta \delta \beta$ patterns, we initialize our extended simulations with five periods each of $\alpha \beta \delta$ and $\alpha \beta \delta \beta$ patterns with the arrangement $[\alpha \beta \delta]_5[\alpha \beta \delta \beta]_5$. 
We begin the first set of simulations with all the solid-solid interfaces having the same interfacial energy of 0.333, and equal liquid diffusivities ($D^l = (1, 1)$). 
Both the patterns are initialized with the same spacing of $\lambda = 270 = 1.33\lambda_{min}^{\alpha \beta \delta} = 1.05\lambda_{min}^{\alpha \beta \delta \beta}$, which corresponds to the undercooling of the $\alpha \beta \delta \beta$ pattern being higher than that of the $\alpha \beta \delta$ pattern ($\Delta T^{\alpha \beta \delta \beta}_{\lambda} > \Delta T^{\alpha \beta \delta}_\lambda$). 
We observe the $\alpha \beta \delta \beta$ pattern transforming to an $\alpha \beta \delta$ pattern, through termination of every alternate $\beta$ phase as shown in Figure \ref{fig:D1_config_2_t_18L}, also called as the short-wavelength  instability in \cite{choudhury2011theoretical}. 
This observation is made for spacings lower than $1.09\lambda_{min}^{\alpha \beta \delta \beta}$. 
The entire morphology transforms to 10 periods of the $\alpha \beta \delta$ pattern of average spacing of $\lambda = 270 = 1.33\lambda_{min}^{\alpha \beta \delta}$ (Figure \ref{fig:D1_config_2_t_60L}), resulting in a reduced average interfacial undercooling compared to the initial configuration with the $\alpha \beta \delta$ and $\alpha \beta \delta \beta$ patterns existing together.  
When initializing the system with a larger spacing ($\lambda = 360$) such that $\Delta T^{\alpha \beta \delta}_\lambda > \Delta T^{\alpha \beta \delta \beta}_\lambda$, the patterns coexist at the same interfacial undercooling by adjusting their average steady state spacing to $\lambda^{\alpha \beta \delta} = 327.1 = 1.62\lambda^{\alpha \beta \delta}_{min}$ and $\lambda^{\alpha \beta \delta \beta} = 383.7 = 1.49\lambda^{\alpha \beta \delta \beta}_{min}$ as shown in Figure \ref{fig:D1_config_2_1204_no_transform}.
Initializing with even higher spacings leads to the onset of oscillatory instabilities in the $\alpha \beta \delta$ phase, and in this paper we limit the study to spacings below the oscillatory regime.

Thus we see in an ideal symmetric ternary alloy with equal interfacial energies and liquid diffusivities, the transformation from $\alpha \beta \delta$ $\to$ $\alpha \beta \delta \beta$ is not observed. 
One of the ways to influence this transformation is by changing the interfacial energies of the $\alpha-\delta$
interface that is absent in the $\alpha\beta\delta\beta$
pattern. In the following, we increase the interfacial energy of the $\alpha-\delta$ interface $\gamma^{\alpha\delta}=0.49$, which essentially increases the undercooling of the $\alpha$ and $\beta$ phases that share an interface with the $\delta$ interface as a result of increased curvature of the respective solid-liquid interfaces.  
For a spacing of $\lambda = 360$, which  corresponds to $\Delta T^{\alpha \beta \delta} > \Delta T^{\alpha \beta \delta \beta}$ with $\gamma^{\alpha \delta} = 0.49$ (Figure \ref{fig:diff_eq}), we do not see a transformation from the $\alpha \beta \delta$ pattern to the expected $\alpha \beta \delta \beta$ pattern. 
The morphology maintains the initial pattern configuration with just an adjustment of the average spacings ($\lambda^{\alpha \beta \delta} = 317.5  = 1.57\lambda^{\alpha \beta \delta}_{min}$ and $\lambda^{\alpha \beta \delta \beta} = 388.3 = 1.51\lambda^{\alpha \beta \delta \beta}_{min}$) leading to an almost equal undercooling for both the patterns (see Figure \ref{fig:D1_config_2_alpha_0p56_1204_no_transform}). 
Moreover, the spacing at which the $\alpha \beta \delta \beta$ pattern transforms to $\alpha \beta \delta$ pattern still remains the same, which is an indication that the 
dynamics of the $\alpha\beta\delta\beta$ morphology is unaffected by the change in the interfacial energy of the interface that does not exist in the pattern.
Thus, we conclude that the morphology does not always result in the lowest undercooling pattern \cite{apel2004lamellar}, and a short-wavelength elimination instability \cite{choudhury2011theoretical} results in the transformation of the $\alpha \beta \delta \beta$ pattern to the $\alpha \beta \delta$ pattern, even though the resulting interfacial undercooling might be higher. 
Hence, interfacial undercooling does not serve as a true indicator of the stability between different pattern arrangements.

To investigate this further, we perform simulations with equal interfacial energies but with unequal diffusivities in the liquid, $D^l = (2, 1)$, thus making the $\beta$ phase richer in the slower diffusing component and resulting in  $\Delta T^{\alpha \beta \delta}_{min} > \Delta T^{\alpha \beta \delta \beta}_{min}$ (Figure \ref{fig:diff_uneq}). 
\begin{figure}[htbp!]
  \centering
    \includegraphics[width=0.45\linewidth]{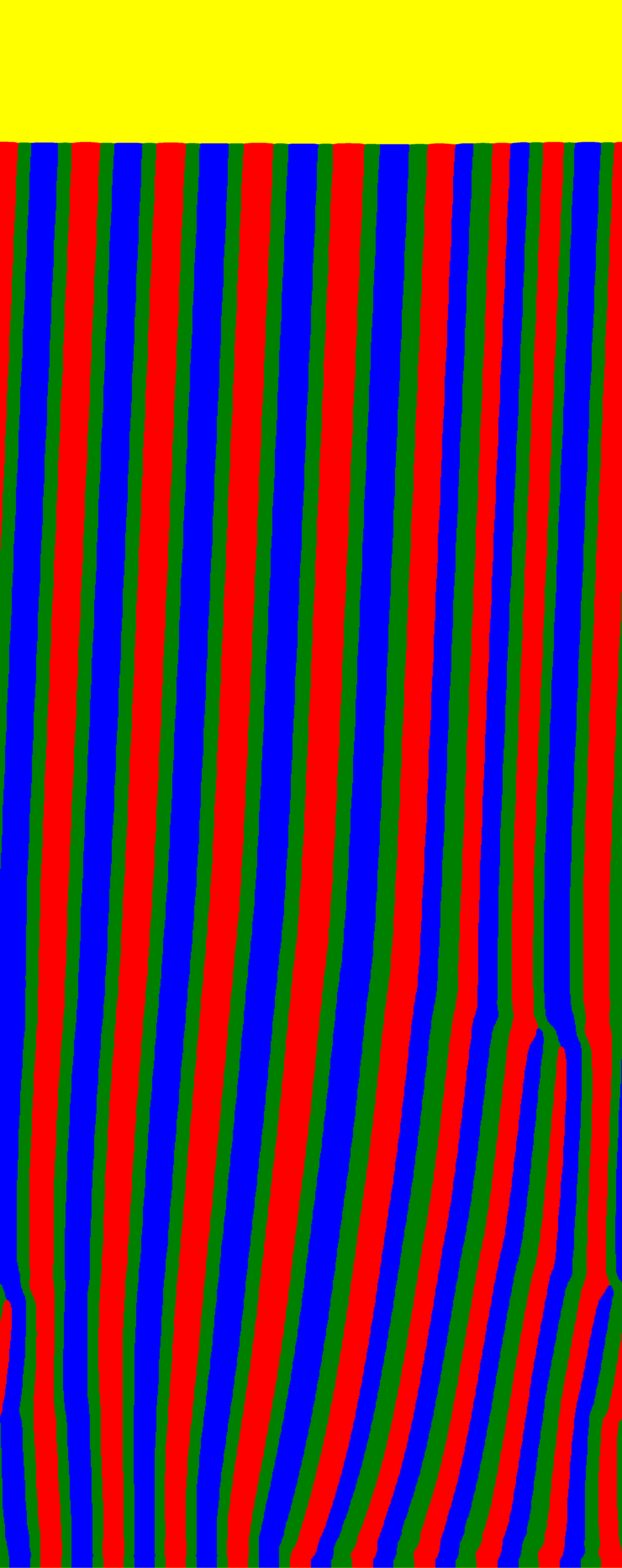}
    \caption{Entire solidification morphology for $D^l = (2, 1)$ starting with an initial configuration $[\alpha\beta\delta\beta]_5[\alpha\beta\delta]_5$ of spacing  $\lambda = 306$ for both the $\alpha\beta\delta$ and $\alpha\beta\delta\beta$ patterns. The steady state morphology has 6 $\alpha\beta\delta\beta$  and 2 $\alpha\beta\delta$ periods. To resolve the features, the image is scaled 3X in the width, hence the tilt looks exaggerated.}
    \label{fig:D2_c2_1024_big}
   \end{figure}
\begin{figure}[htbp!]
  \centering
  \begin{subfigure}[b]{0.3\linewidth}
  \centering
    \includegraphics[width=0.4\linewidth]{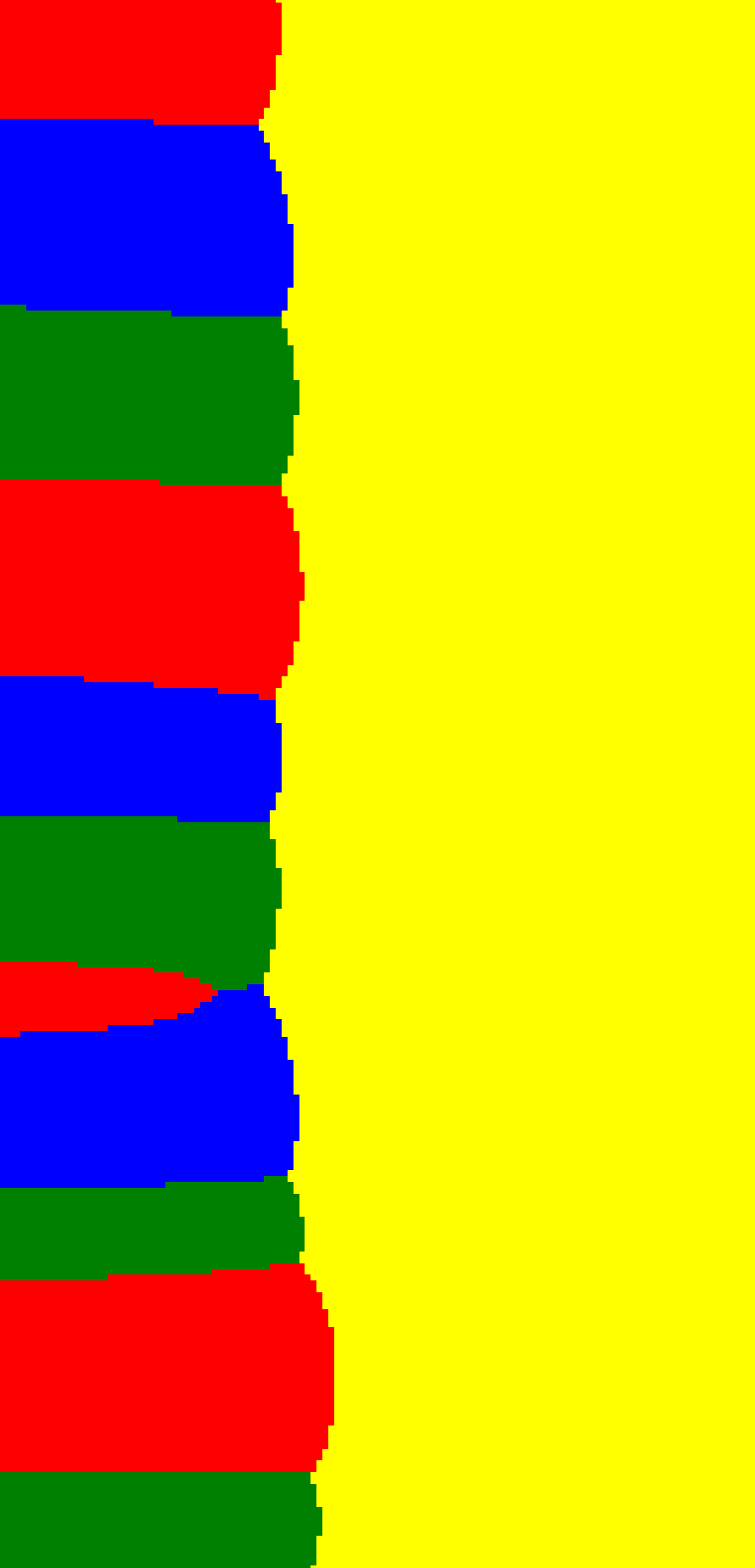}
    \caption{}
    \label{fig:D2_config_2_t_54p5L}
  \end{subfigure}
   \begin{subfigure}[b]{0.3\linewidth}
   \centering
    \includegraphics[width=0.4\linewidth]{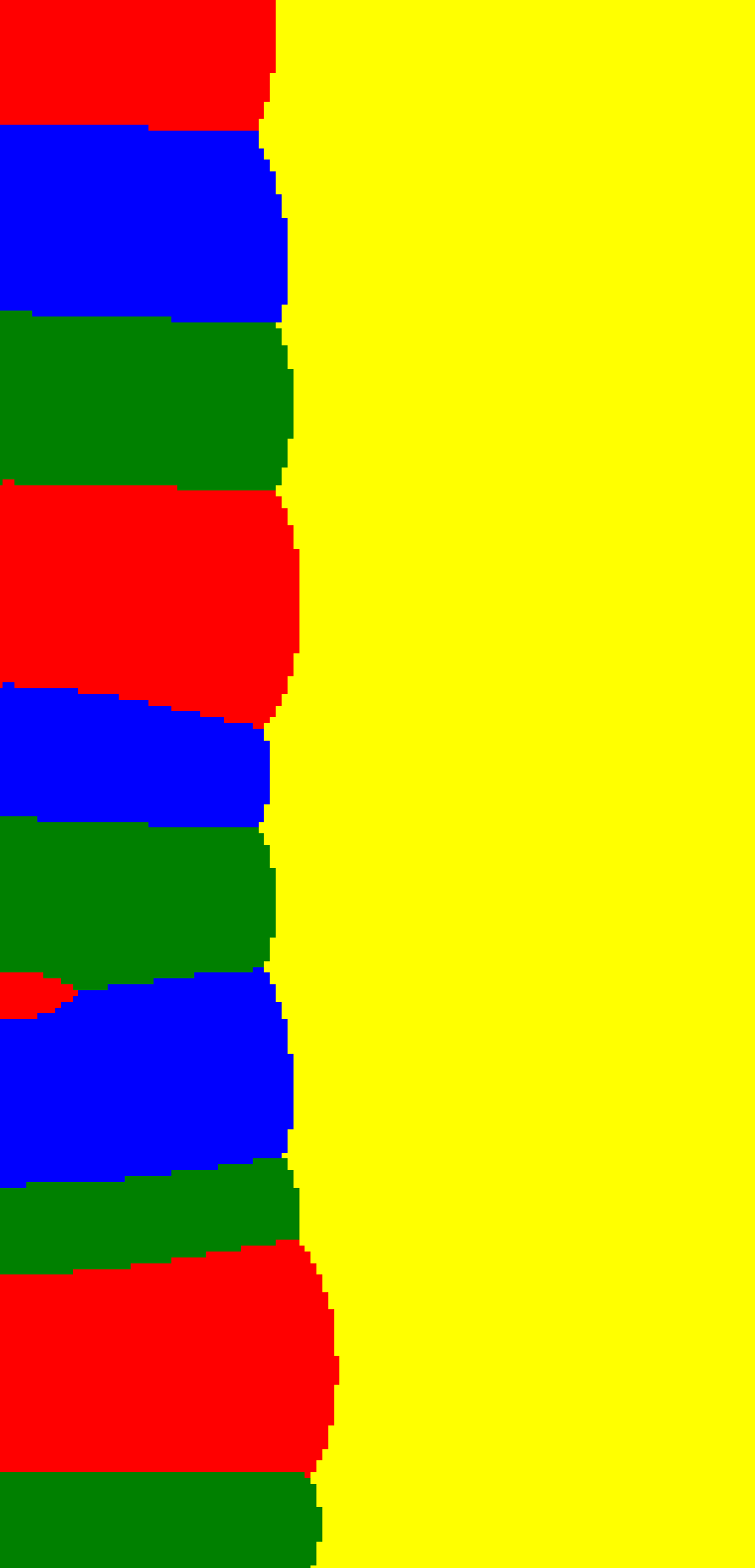}
    \caption{}
    \label{fig:D2_config_2_t_55L}
  \end{subfigure}
     \begin{subfigure}[b]{0.3\linewidth}
     \centering
    \includegraphics[width=0.4\linewidth]{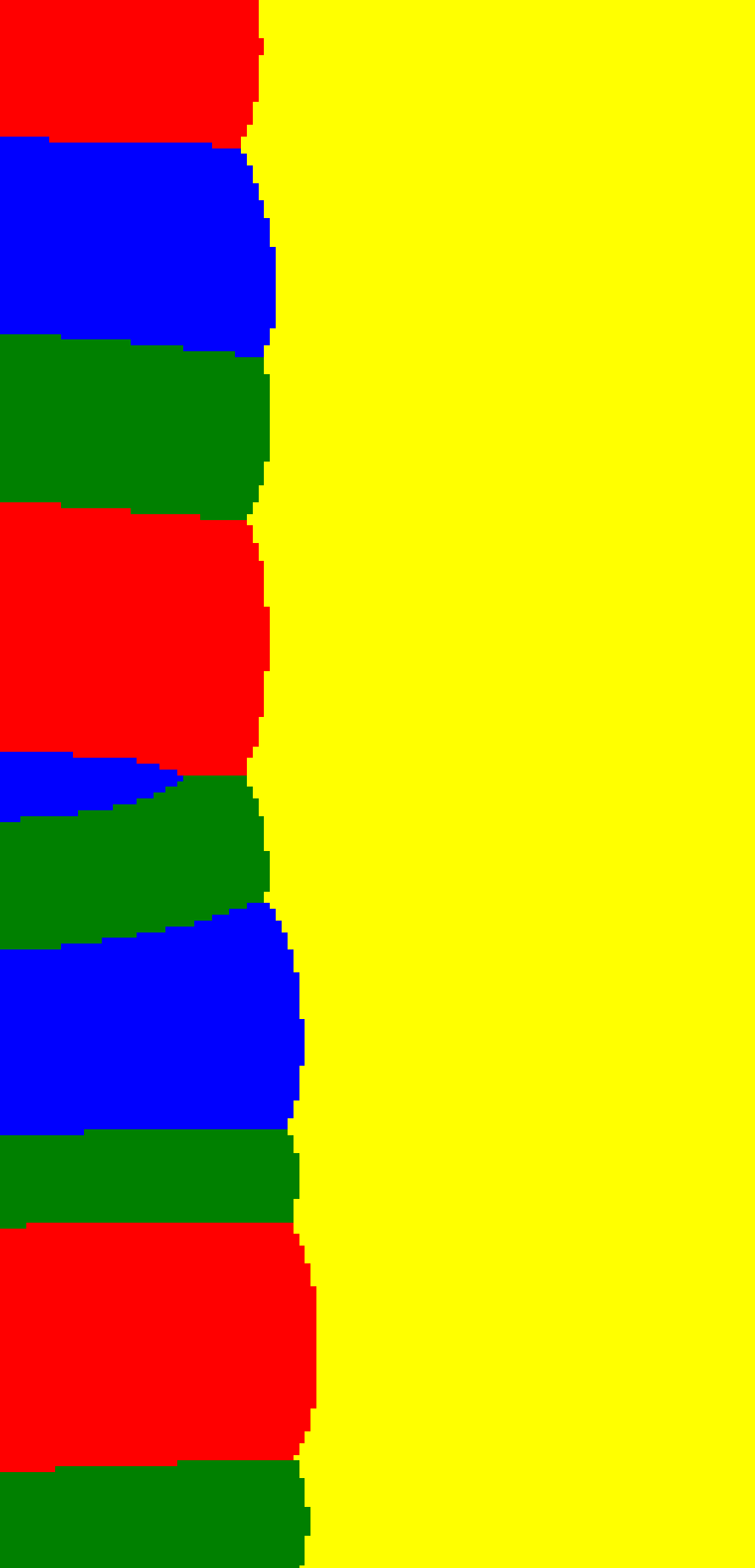}
    \caption{}
    \label{fig:D2_config_2_t_56p5L}
  \end{subfigure}
      \caption{(a, b, c) Zoomed in section showing the time sequence of the $\alpha \beta \delta$ pattern transforming to $\alpha \beta \delta \beta$
      pattern by elimination of $\alpha$ phase lamella followed by $\delta$ phase lamella on either side of the $\beta$ phase lamella for the case of $D^l = (2, 1)$ and initial $\lambda = 306$.}
  \label{fig:D2_config_2}
\end{figure}
We conduct extended simulations with the same initial two pattern configuration for a range of lamellar spacings. For a spacing of $\lambda = 306 = 1.32\lambda_{min,D_2}^{\alpha \beta \delta} = 0.96\lambda_{min,D_2}^{\alpha \beta \delta \beta}$, which corresponds to a higher undercooling of the $\alpha \beta \delta$ pattern with respect to the $\alpha \beta \delta \beta$ pattern ($\Delta T^{\alpha \beta \delta}_{\lambda} > \Delta T^{\alpha \beta \delta \beta}_{\lambda}$), we observe a transformation from the $\alpha \beta \delta$ to the $\alpha \beta \delta \beta$ pattern as shown in the entire solidification morphology in Figure \ref{fig:D2_c2_1024_big}. 
In the initial transient, the $\alpha \beta \delta$ patterns have oscillations that gradually die down.
As also observed in constrained simulations of the $\alpha \beta \delta$ pattern for unequal diffusivities, the $\alpha \beta \delta$ patterns grow with a tilt with respect to the temperature gradient direction, and this also influences the growth of the $\alpha \beta \delta \beta$ patterns as well, that had initially started with a $\ang{0}$ tilt angle.
This tilt occurs in the $\alpha\beta\delta\beta$ pattern due to the modification of the composition profiles in the liquid ahead of the solidification front brought about by the adjoining $\alpha\beta\delta$ patterns. The asymmetry in the composition profiles ahead of the tilted $\alpha\beta\delta$ pattern influences the composition distribution in the liquid ahead of  $\alpha\beta\delta\beta$ pattern resulting in a asymmetry that leads to a tilt. Thus, even a pattern that has mirror symmetry axes bisecting the $\alpha$ and $\delta$ phase lamellae inherits a tilted growth form from an adjoining pattern through the interaction of the diffusion fields.  

Unlike in the previous case where the transformation $\alpha \beta \delta \beta$ $\to$ $\alpha \beta \delta$ occurred by elimination of alternate $\beta$ lamellae, and did not require large spacing adjustments, here the reverse transformation from $\alpha \beta \delta$ $\to$ $\alpha \beta \delta \beta$ progresses through a cascading effect as shown in the transformation events in Figure \ref{fig:D2_c2_1024_big}, and also in the time sequence  in Figure \ref{fig:D2_config_2}. 
The transformation occurs at an $\alpha \beta \delta$ pattern growing with a tilt which has an $\alpha \beta \delta \beta$ pattern as a neighbor, by elimination of $\alpha$ phase lamella followed by $\delta$ phase lamella on either side of the $\beta$ phase lamella.
This mechanism of $\alpha \beta \delta$ transforming to $\alpha \beta \delta \beta$ is also observed experimentally in the In-Bi-Sn eutectic system in \cite{bottin2016stability}.
The steady state pattern has about 6 periods of the $\alpha \beta \delta \beta$ patterns and 2 periods of $\alpha \beta \delta$ pattern which are unable to transform and is possibly a result of the finite simulation size and the number of periods under consideration. The tilt of the morphology is a function of the number of $\alpha\beta\delta$ periods, where the final average tilt angle ($\ang{0.76}$) has reduced compared to the initial average tilt angle ($\ang{2.28}$) when there were a higher number of $\alpha \beta \delta$ periods. The resulting spacing homogenization takes 
longer times to complete with the tilted morphology, where even in same configuration, the spacing is non-uniform and continues to evolve with  time for the duration of the simulation. The average spacing for each pattern is $\lambda_{\alpha \beta \delta \beta} = 420.5 = 1.31\lambda_{min,D_2}^{\alpha \beta \delta \beta}$ and $\lambda_{\alpha \beta \delta} = 268.5 = 1.16\lambda_{min,D_2}^{\alpha \beta \delta}$. 
Due to the non-uniformity of the spacings, the patterns also do not reach a common interfacial undercooling as noticed by the difference in the solid-liquid interfacial position. Thus we see that the presence of unequal diffusivities provides a topological pathway for the transformation from $\alpha \beta \delta$ $\to$ $\alpha \beta \delta \beta$ which was unavailable for the case of equal diffusivities. Here, we 
believe that it is the tilted growth form of the $\alpha\beta\delta$ pattern that is central to this transformation. The variation of the tilt is influenced by the magnitude of the solutal diffusivity contrast as highlighted in Figure \ref{fig:D2_angle_ABC} as well as the spacings. The magnitude of the tilt determines the possibility of this transition from the simpler to the more complicated pattern, with a higher tilt leading to an increased possibility for the transformation. 

While these simulations involving configurations of patterns that are composed of multiple periods of simpler patterns  provide insights about transformation mechanisms between configurations, in general, the patterns arising in experiments need not correspond to just these simpler patterns.
In the following, we investigate whether there exists any strong tendency for pattern selection in response to the changes in the solid-solid interfacial energy or the solutal diffusivities.

\subsubsection{Complex patterns}
\label{sec:complex}
\begin{figure}[htbp!]
  \centering
  \begin{subfigure}[b]{0.9\linewidth}
  \centering
    \includegraphics[width=0.9\linewidth]{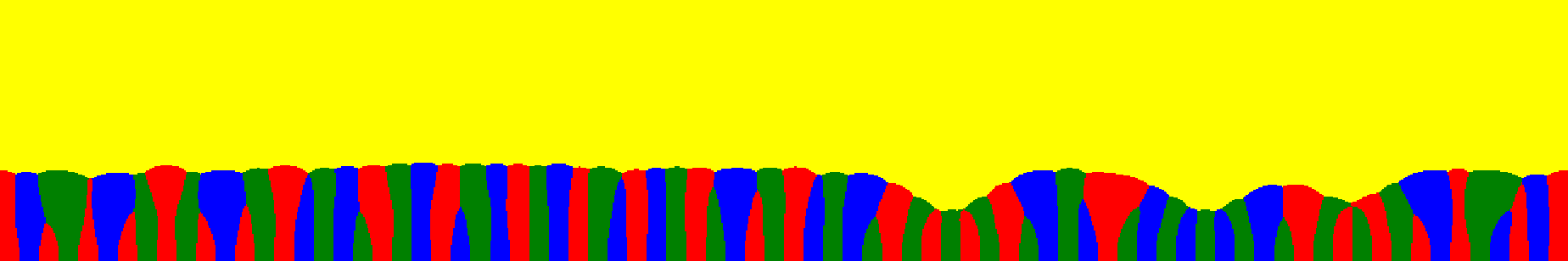}
    \caption{}
    \label{fig:D1_config_4_t_100000}
  \end{subfigure}
  
   \begin{subfigure}[b]{0.9\linewidth}
   \centering
    \includegraphics[width=0.9\linewidth]{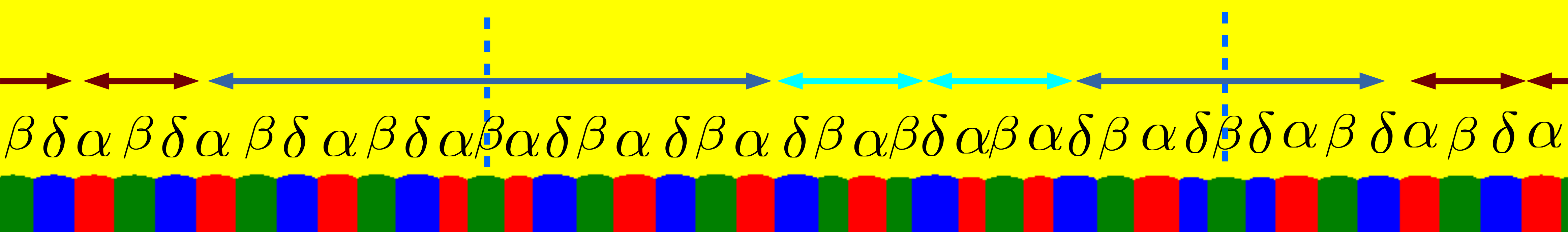}
    \caption{}
    \label{fig:D1_config_4_t_9000000}
  \end{subfigure}
  
   \begin{subfigure}[b]{0.9\linewidth}
   \centering
    \includegraphics[width=0.9\linewidth]{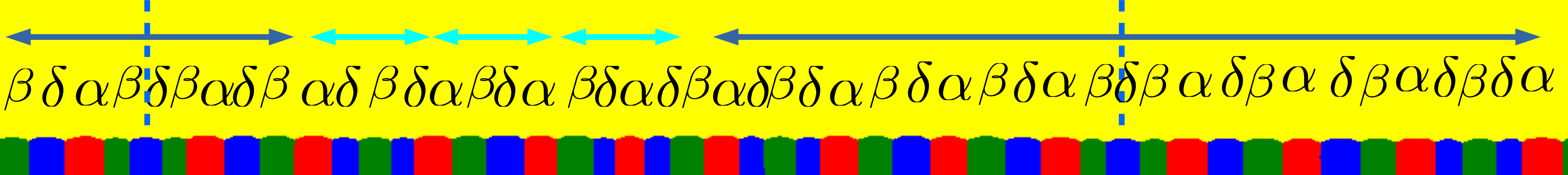}
    \caption{}
    \label{fig:D1_config_4_t_6000000_my_1804_dx_2}
  \end{subfigure}
  
  \begin{subfigure}[b]{0.9\linewidth}
  \centering
    \includegraphics[width=0.9\linewidth]{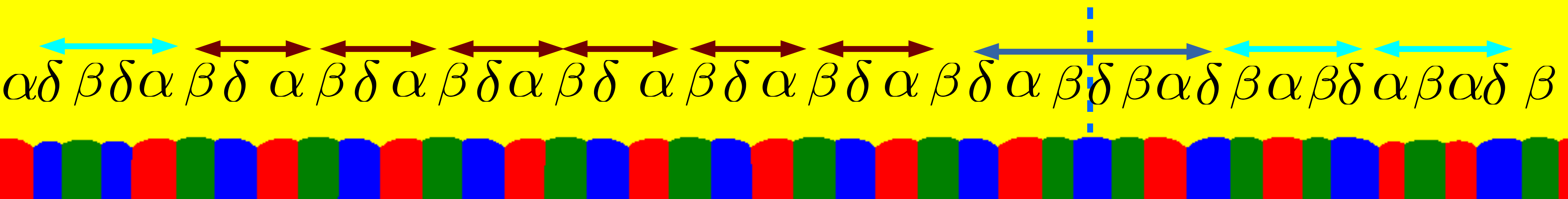}
    \caption{}
    \label{fig:D1_config_4_s_0p49_t_9300000_my_1204}
  \end{subfigure}
  
    \caption{(a) Initial random arrangement of phases showing merging and termination of lamellae for $D^l = (1, 1)$ and all the solid-solid interfacial energies are 0.333. (b, c, d) Steady state configurations. Existence of long mirror symmetric configurations is observed (spanned by the blue arrow). The 3 cycle patterns are represented by the brown arrows, and the 4 cycle patterns by the cyan arrow.
    (b) The steady state morphology when all the solid-solid interfacial energies are 0.333 is represented as $[\alpha\beta\delta]_3 \overline{[\alpha\beta\delta]_2[\alpha\beta\alpha][\delta\beta\alpha]_2} [\delta\beta\alpha\beta]_2 \overline{[\delta\beta\alpha][\delta\beta\delta][\alpha\beta\delta]}$, where the overline denotes a mirror symmetric pattern.
    (c) On reducing the initial random lamellae size, the steady state configuration is $\overline{[\beta\delta\alpha][\beta\delta\beta][\alpha\delta\beta]} [\alpha\delta\beta\delta] [\alpha\beta\delta\alpha] [\beta\delta\alpha\beta] \overline{[\alpha\delta\beta][\delta\alpha\beta]_3 \delta [\beta\alpha\delta]_3[\beta\alpha\delta]}$.
    (d) On increasing $\gamma^{\alpha \delta}$ to 0.49, the steady state morphology is  $[\beta\delta\alpha]_6\overline{[\delta\alpha\beta]\delta[\beta\alpha\delta]} [\beta\alpha\beta\delta][\alpha\beta\alpha\delta] [\beta\alpha] [\delta\beta\delta\alpha]$.}
  \label{fig:D1_config_4}
\end{figure}

We assess pattern selection by performing simulations
with a random arrangement of a large number of solid phase lamellae (nuclei) of small width. This demands that the spacings of the patterns increase by the termination and merging of lamellae, and finally evolve into a steady state pattern. 
Just like the previous studies on pattern competition, here as well we conduct a number of simulations by varying the solid-solid interfacial energies and liquid diffusivities, but starting from a random initial configuration.
For equal interfacial energies and equal liquid diffusivities ($D^l = (1, 1)$), Figure \ref{fig:D1_config_4_t_100000} shows the initial transient where the lamellae increase their spacing by termination and merging events, and the steady state profile is shown in Figure \ref{fig:D1_config_4_t_9000000}. 
Unlike the previous extended case with equal diffusivities and interfacial energies having initial configurations of the type $[{\alpha \beta \delta \beta}]_{5}[{\alpha \beta \delta}]_{5}$, where the steady state patterns depended upon the initial spacing and were predominantly composed of simple patterns of the type $\alpha\beta\delta$ or $\alpha\beta\delta\beta$, here in addition to the simple patterns, we observe long mirror symmetric complicated patterns of the form $[\alpha\beta\delta]_2[\alpha\beta\alpha][\delta\beta\alpha]_2$ and $[\delta\beta\alpha][\delta\beta\delta][\alpha\beta\delta]$ in Figure \ref{fig:D1_config_4_t_9000000} where we have factorized the complicated patterns into periods of the simplest cycles. 
In order to understand the influence of the size of the initial lamellae on the final pattern, we conduct another simulation with the same setup as described above, but with a lower initial lamellar size, and thus a higher number of initial lamellae.
We again observe long mirror symmetric patterns of the type $[\alpha\delta\beta][\delta\alpha\beta]_3 \delta [\beta\alpha\delta]_3[\beta\alpha\delta]$ (Figure \ref{fig:D1_config_4_t_6000000_my_1804_dx_2}), along with an increase in the total number of lamellae which is a consequence of the reduced initial random lamellae size. Moreover, upon increasing $\gamma^{\alpha \delta}$ to $0.49$, similar patterns as in the above cases are observed which consist of periods of simple 3 and 4 cycle patterns ($[\beta\delta\alpha]_6$) along with a complicated mirror symmetric pattern ($[\delta\alpha\beta]\delta[\beta\alpha\delta]$) (Figure \ref{fig:D1_config_4_s_0p49_t_9300000_my_1204}). 
From these simulations starting with a random initial configuration, we see that in the steady state morphology, simple 3 and 4 cycle patterns along with long mirror symmetric patterns are selected for $D^l = (1, 1)$ and the changes in the solid-solid interfacial energies extends a weak influence on pattern selection, although there is strong influence on the positions of the undercooling vs. spacing curves as revealed in Figure \ref{fig:diff_eq}.
However, this result ignores the influence that the solid-solid interfacial energy would have on the distribution of nuclei at the start of solidification in actual experimental conditions.

\begin{figure}[htbp!]
  \centering
    \includegraphics[width=0.5\linewidth]{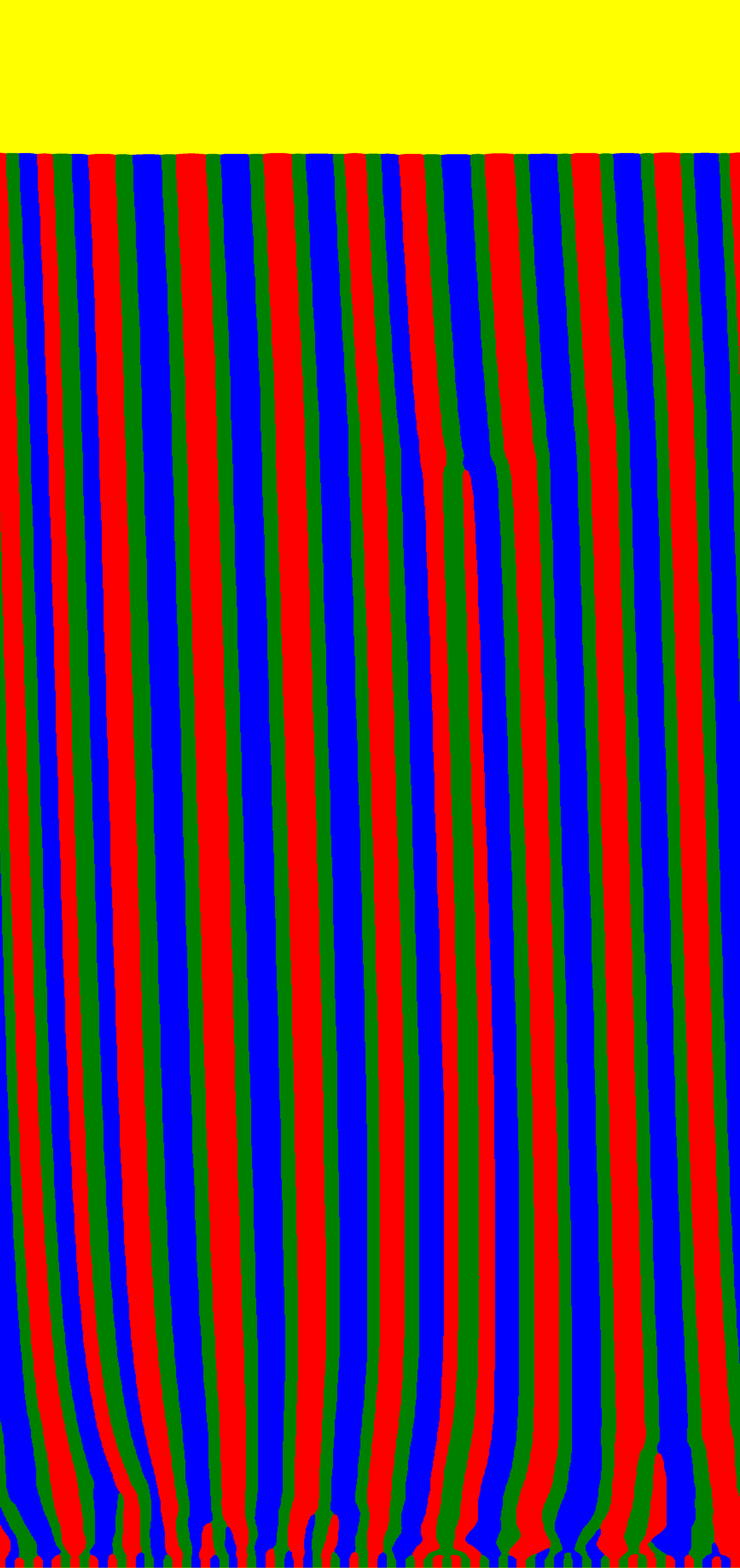}
    \caption{Entire solidification morphology starting with a random initial configuration for $D^l = (2, 1)$ and all interfacial energies of 0.333. The steady state morphology has predominantly $\alpha \beta \delta \beta$ patterns. To resolve the features, the image is scaled 3X in the width.}
    \label{fig:D2_c2_1024_big_random}
   \end{figure}

Figure \ref{fig:D2_c2_1024_big_random} shows the morphology obtained when $D^l = (2, 1)$ and all solid-solid interfacial energies are 0.333. 
The initial transient shows elimination and merging events resulting in an initial selection of predominantly $\alpha \beta \delta \beta$ patterns. We also observe the transformation of the $\alpha \beta \alpha \delta$ pattern to the $\alpha \beta \delta \beta$ pattern after a long time into the solidification regime that involves the elimination of a single $\alpha$ lamella (note that $\Delta T^{\alpha\beta\delta\beta} < \Delta^{\alpha\beta\alpha\delta}$).
The steady state morphology is predominantly $\alpha \beta \delta \beta$ (7 periods), with 2 $\alpha \beta \delta$ periods, having an average spacing of 397 ($1.24\lambda_{min,D_2}^{\alpha \beta \delta \beta}$) and 275 ($1.18\lambda_{min,D_2}^{\alpha \beta \delta}$) respectively, and a steady state average tilt angle of $\ang{1}$. 
Thus we have obtained a morphology that majorly contains the $\alpha \beta \delta \beta$ pattern which has the lowest minimum undercooling for $D^l = (2, 1)$. This result can be seen as an influence of the lowered constitutional undercooling in front of the $\beta-l$ interface in the $\alpha\beta\delta\beta$ configurations (see Figure \ref{fig:undercooling_constitutional_beta}) that allows such pattern occurrences to grow ahead than others. 
Interestingly, since the relative diffusivity of the component $B$ is lower, it also leads to a steeper diffusion gradient corresponding to the component $B$ ahead of the $\beta$ phase. The corresponding microstructural size of the $\beta$ phases that result in the final pattern bears a reflection to the reduced diffusion distances of the component $B$ ahead of the $\beta$ solidification front, where due to the larger frequency of $\beta$ lamellae, the corresponding widths are lower than the lamellae of the other phases.

\begin{figure}[htbp!]
  \centering
    \includegraphics[width=0.9\linewidth]{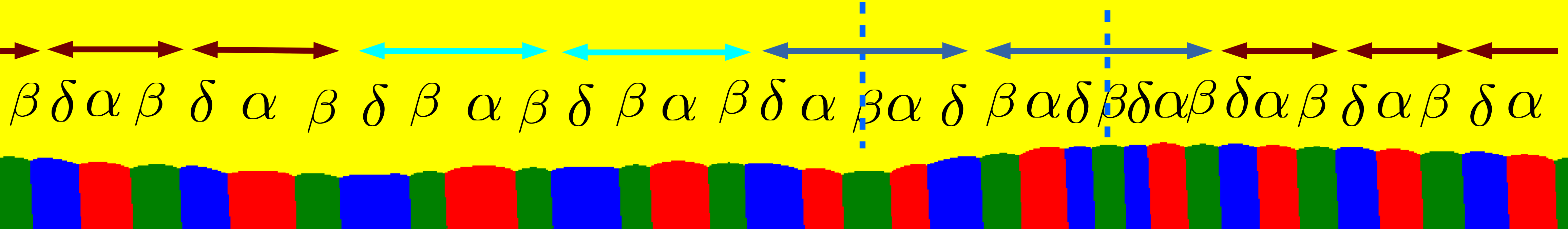}
    \caption{Steady state morphology for $D^l = (2, 1)$ starting with a random configuration with reduced $\gamma^{\alpha \delta} = 0.24$. 
    The morphology has the configuration $[\delta\alpha\beta]_5 [\delta\beta\alpha\beta]_2 \overline{[\delta\alpha\beta\alpha\delta]}\;\overline{[\beta\alpha\delta][\beta][\delta\alpha\beta]}$, where the overline denotes a mirror symmetric pattern. Brown arrows span $\alpha\beta\delta$ patterns and cyan arrows span $\alpha\beta\delta\beta$ patterns.
    }
  \label{fig:D2_config_4}
\end{figure}

\begin{figure}[htbp!]
  \centering
    \includegraphics[width=0.9\linewidth]{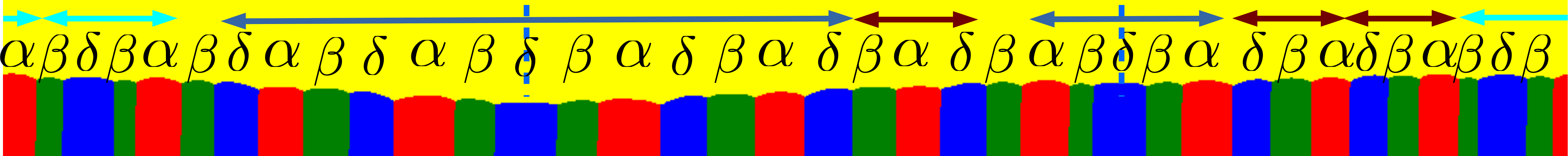}
  \caption{Steady state morphology for $D^l = (1.6, 1)$. The morphology has the configuration $[\beta\delta\beta\alpha]_2 \beta \overline{[\delta\alpha\beta]_2 \delta [\beta\alpha\delta]_2}[\beta\alpha\delta\beta][\overline{\alpha\beta\delta\beta\alpha]} [\delta\beta\alpha]_2$, where the overline denotes mirror symmetric patterns (spanned by the blue arrow in the Figure). Brown arrows span $\alpha\beta\delta$ patterns and cyan arrows span $\alpha\beta\delta\beta$ patterns.}
  \label{fig:D1p6_config_4}
\end{figure}

Next, we reduce $\gamma^{\alpha \delta}$ to $0.24$ while retaining the diffusivity matrix $D^l = (2, 1)$, thus decreasing the interfacial undercooling of the $\alpha \beta \delta$ pattern as compared to the previous case with all interfacial energies equal to 0.333, and the final morphology is shown in Figure \ref{fig:D2_config_4}. Unlike the previous case of $D^l = (2, 1)$ and equal interfacial energies, here we see that  there is no clear winner and the final configuration shows the presence of 3 and 4 cycle patterns along with complicated mirror symmetric patterns ($[\delta\alpha\beta\alpha\delta]$ and $[\beta\alpha\delta][\beta][\delta\alpha\beta]$), although the length of the mirror symmetric patterns has reduced compared to those observed with equal diffusivities having equal interfacial energies. Due to the reduced length of mirror symmetric patterns and highest diffusivity contrast ($D^l = (2, 1)$), this configuration also exhibits the largest tilt angle of  all the extended simulations of $\ang{2.65}$.
A similar situation arises for a simulation with $D^l = (1.6, 1)$ and equal interfacial energies, for which the $\alpha \beta \delta$ and $\alpha \beta \delta \beta$ patterns have nearly equal values of $\Delta T^{\alpha \beta \delta}_{min} = \Delta T^{\alpha \beta \delta \beta}_{min}$ (Figure \ref{fig:diff_uneq_1p6}). 
The steady state configuration is again complex (Figure \ref{fig:D1p6_config_4}), with no clear winner, 
having the presence of mirror symmetric segments ($[\delta\alpha\beta]_2 \delta [\beta\alpha\delta]_2$) along with the $\alpha\beta\delta$ pattern, and an average tilt angle of $\ang{0.72}$. 

In summary, there is no evidence from our simulations to suggest that there is a strong selection of a given pattern in response to the change in the solid-solid interfacial energy or the solutal diffusivities, when starting from random configurations. The pattern that emerges is a result of the complex interplay of the availability of transition mechanisms between configurations, instabilities and the asymmetry in the growth morphologies.
In this light, one does not expect to find a selection among the patterns. 
In particular, the only case we find a near clear winner is for the simulation in Figure \ref{fig:D2_c2_1024_big_random} where we find a strong influence of the diffusivity contrast on the selection of the pattern $\alpha\beta\delta\beta$. The length scale of the lamella corresponding to the phase richer in the slower diffusing species is finer compared to the width of the lamella of the other phases, which is a reflection of the lower diffusion length of the slower diffusing species in the liquid, ahead of the phase richer in the slower diffusing species. A key factor that possibly influences the appearance of this pattern, is that the short-wavelength instability that limits the spacings over which such patterns remain stable is the lowest for this case ($\lambda_{el}^{sw}=0.82\lambda_{min, D_2}^{\alpha\beta\delta\beta}$) and thereby this pattern has the largest range of stable spacings when $D^l=(2,1)$ and interfacial energies are equal. However, the situation becomes complex when there is a superposition of the asymmetries in the diffusivity and the interfacial energies. In general, we find that the introduction of any asymmetry reduces the length of the mirror symmetric patterns that appear in the final microstructure. Additionally, among the asymmetries, a key differentiation is that, contrast in the solutal diffusivity also leads to tilted patterns that is brought about by the associated asymmetry in the composition field ahead of the solid-liquid interface. This tilt in the morphology that typically occurs for patterns without a mirror symmetric axes, is also transmitted to the co-existing mirror symmetric patterns through the interaction of the composition fields in the liquid, thus giving rise to a globally tilted morphology with respect to the imposed temperature gradient.

\section{Conclusion}
In this paper, we conduct constrained and extended phase-field simulation studies  of thin-film directional solidification in a model symmetric three-phase eutectic alloy in order to investigate pattern formation influenced by changes in solid-solid interfacial energies and diffusivities.
We begin with constrained simulations in order to obtain insights about the solid-liquid interfacial undercooling vs. spacing variations for the simplest patterns possible, of cycle lengths 3 and 4 under different conditions of solid-solid interfacial energies and liquid diffusivities. Expectedly, we find that the changes in the solid-solid interfacial energies influence the position of the undercooling curves through a modification of the curvature undercooling, while the introduction of contrast in solute diffusivities modifies the constitutional undercooling. In particular,  we find that when the relative diffusivity of component B is lowered, the minimum undercooling of the $\alpha\beta\delta\beta$ pattern is the lowest where the $\beta$ phase is richer in the slower diffusing species. With respect to the morphology, a contrast in the solute diffusivity leads to a tilt with respect to the temperature gradient, where the extent of the tilt scales with the magnitude of diffusivity mismatch as well as the spacings. The tilt arises because of a lack of mirror symmetry in the $\alpha\beta\delta$ pattern in contrast to the $\alpha\beta\delta\beta$ pattern.
Thereafter, we investigated the stability of the patterns to long-wavelength perturbations. Here, we observed that in contrast to the Eckhaus type instability leading to lamella elimination that sets the lower bound of spacings for the $\alpha\beta\delta$ pattern, for the 
$\alpha\beta\delta\beta$ pattern it is the short-wavelength instability
\cite{choudhury2011theoretical}, that occurs before the Eckhaus instability for the case of equal diffusivities and interfacial energies. This transition that involves the increase in width of one of the smaller $\beta$ lamella and reduction in the other, in a given $\alpha\beta\delta\beta$ pattern, becomes unstable for spacings below the short-wavelength stability limit and leads to the transition to the simpler $\alpha\beta\delta$ morphology. The position of the short-wavelength and Eckhaus instability limits determines the temporal evolution to long-wavelength perturbations at small spacings. The relative positions are a function of the solute diffusivity contrast, where we find that on reducing the relative diffusivity of the species $B$, the short-wavelength instability limit is lowered for the $\alpha\beta\delta\beta$ where the $\beta$ phase is richer in the slower diffusing species. Subsequently, we have investigated pattern competition between the two simplest patterns as a function of the asymmetry in interfacial energy and the solute diffusivity. Here, we find that the co-existence of the simplest patterns are a function of the spacings and diffusivity contrast. For equal diffusivities, the simplest patterns can co-exist when the spacing of the $\alpha\beta\delta\beta$ pattern is larger than the short-wavelength instability limit and the steady state microstructure of the co-existing morphologies is such that the undercoolings of the two morphologies are nearly equal. On introduction of a solutal diffusivity contrast, the co-existing patterns of $\alpha\beta\delta$ and $\alpha\beta\delta\beta$, develops a tilt because of the interaction between the composition fields ahead of the two patterns, where the asymmetry in composition ahead of the $\alpha\beta\delta$ pattern modifies the otherwise symmetric composition profiles ahead of the $\alpha\beta\alpha\delta$ solidification front, resulting in a tilt in the entire coexisting pattern.
The magnitude of the tilt determines the coexistence of the patterns. For larger diffusivity contrasts that lead to increased tilts, e.g. $D^l=(2,1)$, we find a new transition mechanism that leads to the transformation from the $\alpha\beta\delta$ to the $\alpha\beta\delta\beta$ pattern which involves a coordinated elimination of the $\alpha$ and $\delta$ lamella from the $\alpha\beta\delta$ morphology. This mechanism might explain the more usual observations of mirror symmetric patterns of the $\alpha\beta\delta\beta$ type in experiments. 
Thereafter, we investigate the formation of complex patterns starting from random configurations, where we find the occurrence of large mirror symmetric sequences for the equal interfacial energies and solute diffusivities whose length diminishes on introduction of asymmetry in the solute diffusivity and the interfacial energy.

Finally, while the simulation studies bring out insights about the pattern formation during thin-film growth, the results must be assessed in light of the consideration that in reality experiments are only quasi-2D. The presence of a separation in between the glass plates in actual experiments provides for 3D transformation pathways that are not present in the simulation studies performed in this paper. These transformation pathways include the possibility of phase invasion \cite{bottin2016stability} from behind the viewing plane leading to possibilities for lamellar width refinement as well as the formation of patterns that might not be possible in the present simulation studies. In reality such simulations are rather difficult as they must not only include the physics of the solidification phenomena but also the interaction of the solid and liquid phases with the glass walls. Inclusion of such effects remains a  scope for future work. Further, while we have studied the influence of solid-solid interfacial energies by lowering one of the three possible interfaces, in real alloys the interfaces are also known to be anisotropic. The influence of such anisotropy on pattern selection during thin-film growth is certainly a scope for future simulation studies.  

\section{Acknowledgement}
The authors would like to thank DST-SERB, India for funding through the project (DSTO1679). 
SK would like to thank SERC and TUE-CMS, IISc for providing access to high-performance computational resources, including the use of the SahasraT (Cray XC40) machine at SERC.

\section{Conflict of Interest} 
This study was funded by DST-SERB, India through the project (DSTO1679). The authors declare that they have no conflict of interest.

\section{Data and code availability}
All the data required for the reproduction of the results in the paper are already mentioned in the paper. The phase-field code used for the generation of the results cannot be shared at this point of time, however all the details of the model formulation are present in the paper from which the code can be created.

\section{Ethical approval}
All the authors have read and approved the submission.
The study did not involve any experiment on human or animal subjects.

\typeout{}
\bibliography{symmetric_ternary_2d}

\section{Graphical abstract}
\begin{figure}[htbp!]
  \centering
    \includegraphics[width=0.9\linewidth]{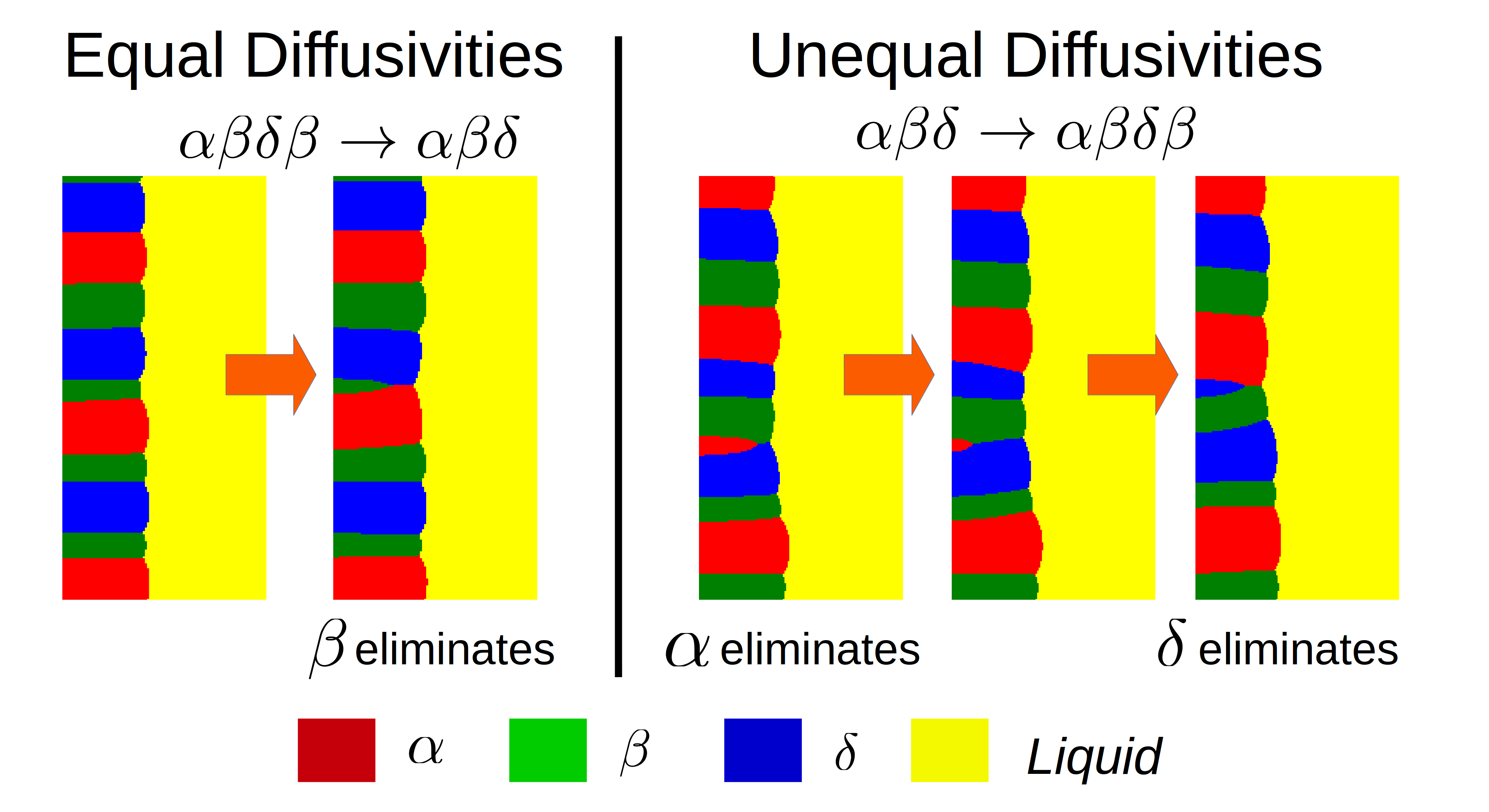}
\end{figure}

\end{document}